\documentclass[prb,twocolumn,superscriptaddress,amsmath,amssymb]{revtex4-2}
\usepackage[colorlinks=true,citecolor=blue,linkcolor=blue,breaklinks=true]{hyperref}

\usepackage{color,graphicx,soul}
\usepackage{bm}
\usepackage{amsmath}
\usepackage{amssymb}
\usepackage{mathrsfs}
\usepackage{newtxmath}
\usepackage{caption}
\usepackage[usenames,dvipsnames]{xcolor}
\usepackage{verbatim}
\usepackage{float}
\usepackage[justification=Justified]{caption}
\usepackage[utf8]{inputenc}

\begin{document}
	
	\title{Rashba spin-orbit coupling and artificially engineered topological superconductors}

    \author{Sankar Das Sarma}
    \affiliation{Condensed Matter Theory Center and Joint Quantum Institute, Department of Physics, University of Maryland, College Park, MD 20742, USA}
	\author{Katharina Laubscher}
    \affiliation{Condensed Matter Theory Center and Joint Quantum Institute, Department of Physics, University of Maryland, College Park, MD 20742, USA}
    \author{Haining Pan}
    \affiliation{Department of Physics and Astronomy, Center for Materials Theory, Rutgers University, Piscataway, NJ 08854 USA}
	\author{Jay D. Sau}
    \affiliation{Condensed Matter Theory Center and Joint Quantum Institute, Department of Physics, University of Maryland, College Park, MD 20742, USA}
    \author{Tudor D. Stanescu}
    \affiliation{Department of Physics and Astronomy, West Virginia University, Morgantown, WV 26506, USA}

	\date{\today}

    \begin{abstract}
	One of the most important physical effects in condensed matter physics is the Rashba spin-orbit coupling (RSOC), introduced in seminal works by Emmanuel Rashba.  In this article, we discuss, describe, and review (providing critical perspectives on) the crucial role of RSOC in the currently active research area of topological quantum computation.  Most, if not all, of the current experimental topological quantum computing platforms use the idea of Majorana zero modes as the qubit ingredient because of their non-Abelian anyonic property of having an intrinsic quantum degeneracy, which enables nonlocal encoding protected by a topological energy gap.  It turns out that RSOC is a crucial ingredient in producing a low-dimensional topological superconductor in the laboratory, and such topological superconductors naturally have isolated localized midgap Majorana zero modes.  In addition, increasing the RSOC strength enhances the topological gap, thus enhancing the topological immunity of the qubits to decoherence.  Thus, Rashba's classic work on SOC may lead not only to the realization of localized non-Abelian anyons, but also fault-tolerant quantum computation.
	\end{abstract}
 
	\maketitle
	
	\section{Introduction}
    \label{sec:intro}

    Emmanuel Rashba was a theoretical giant in physics, and perhaps his most-known work pertains to what is often referred to as the Rashba effect or Rashba spin-orbit (SO) coupling~\cite{rashba1959symmetry,bychkov1984properties}. 
    The effect, which is relativistic in origin, arises from spatial inversion asymmetry (SIA) leading to an effective electronic SO coupling (SOC) causing a momentum-dependent spin splitting in compound semiconductors. Although the fundamental and deep significance of the Rashba effect was not appreciated for a long time, perhaps because the effect could often be quantitatively weak, it is now well-established to be an important part of physics playing a key role in many phenomena~\cite{winkler2003spin,zutic2004spintronics}. 
    The current article discusses the key role played by the Rashba effect in the proposal for creating artificial 2D and 1D structures hosting Majorana zero modes, which can lead to topological quantum computing~\cite{nayak2008nonabelian}.
    The subject is of considerable fundamental significance because the Majorana zero modes are exotic non-Abelian anyons, which may arise as mid-gap localized neutral quasiparticle excitations in topological superconductors. In addition, the subject is of great technological significance because it could lead to a fault-tolerant quantum computer. In fact, Microsoft Corporation has chosen this approach as their quantum computing platform with hundreds of Microsoft scientists working on the realization of this platform, where the Rashba effect is a key ingredient~\cite{microsoftquantum2023inasal,aghaee2025interferometric,aasen2025roadmap}.
    If Microsoft succeeds in creating a fault-tolerant topological quantum computer, the Rashba coupling becomes a centerpiece of a disruptive new technology.

Majorana zero modes (MZMs) arise naturally as `defects' or `localized quasiparticles' at an interface between a low-dimensional topological superconductor and other (non-topological) materials. A classic example is the Kitaev chain model, where a 1D lattice is assumed to be a spinless $p$-wave superconductor with electrons hopping between nearest-neighbor sites~\cite{kitaev2001unpaired}. The model is exactly solvable, and the solution manifests a quantum phase transition between a trivial superconductor and a topological superconductor with the appropriate tuning of the hopping parameter and the chemical potential.  At the transition point, representing a topological quantum phase transition (TQPT), the superconducting gap vanishes with the trivial gap closing on one side and a topological gap opening on the other side. The topological phase is characterized by the bulk topological gap as well as localized MZMs at the wire ends. For a long enough wire, where the wire length is much larger than the superconducting coherence length, the end-localized MZMs are zero-energy non-Abelian anyons, suitable for topological quantum computing. The Kitaev chain idea was, however, just a toy model, demonstrating the possibility of MZMs in a 1D wire with no suggestions for how to realize such a system experimentally. Indeed, Kitaev simply assumed the existence of spinless $p$-wave superconductivity in the 1D chain, but such spinless $p$-wave superconductivity does not seem to exist naturally in any material. The Kitaev chain idea was simply ignored by the community for a decade since it was not a practical proposal for any experimental realization.

\begin{figure*}[tb]
    \centering
    \includegraphics[width=0.9\textwidth]{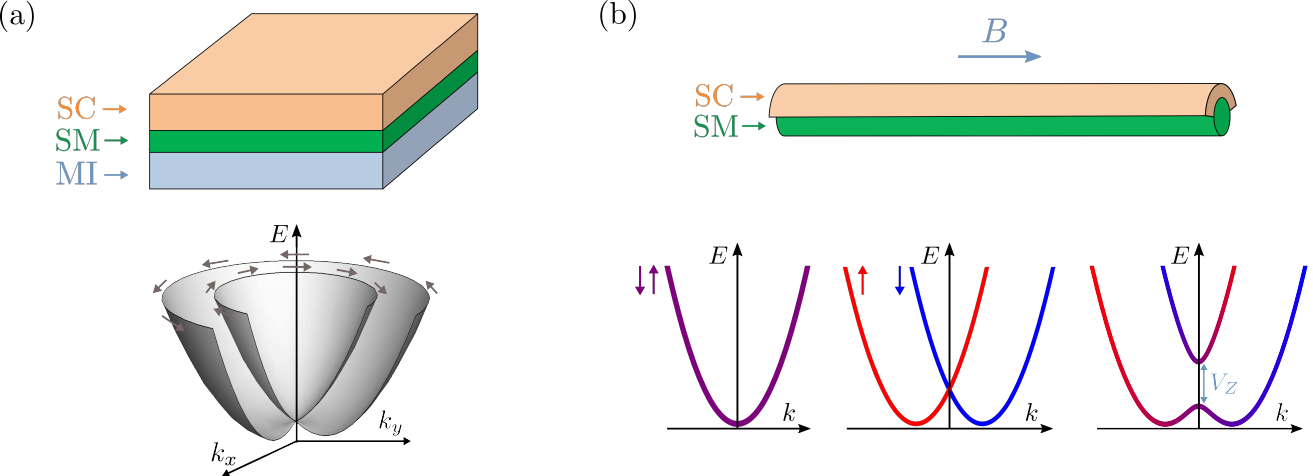}
    \caption{
    (a) Top: Heterostructure composed of a semiconductor (SM) layer sandwiched between an $s$-wave superconductor (SC) and a magnetic insulator (MI). Bottom: RSOC leads to two spin-split Fermi surfaces in the single-particle energy spectrum of the SM layer. In the presence of a finite proximity-induced Zeeman splitting (not shown here), an energy gap is opened in the SM band structure near $(k_x,k_y) = (0,0)$. If the chemical potential lies in this gap, the system has a single Fermi surface similar to a topological insulator. (b) Top: Semiconductor nanowire proximitized by an $s$-wave superconductor with a magnetic field applied along the wire axis. Bottom left: Without SOC, the band structure of the semiconductor nanowire is spin-degenerate. Bottom center: RSOC leads to a relative shift between the spin-up (red) and spin-down (blue) bands in momentum space. Bottom right: A magnetic field applied along the wire axis leads to a Zeeman splitting $V_Z$ that mixes and splits the bands near $k = 0$. If the chemical potential is placed in the Zeeman gap, the system becomes a helical 1D nanowire with a single pair of spin-momentum locked low-energy modes.
    }
    \label{fig:setup}
\end{figure*}

The situation changed in 2009-2010 with the appearance of specific and practical theoretical ideas for the realization of an effective spinless $p$-wave superconductivity using engineered semiconductor-superconductor (SM-SC) hybrid structures, where a proximity-induced spinful $s$-wave superconductivity from a regular SC (e.g., Al) in a semiconductor (e.g., InAs) is transformed into an effective spinless $p$-wave superconductivity by appropriately combining Rashba SOC (RSOC) and spin-splitting (induced, for example, by an applied Zeeman field or an exchange field induced by a nearby ferromagnetic insulator) in the SM~\cite{sau2010generic,sau2010nonabelian,lutchyn2010majorana,alicea2010majorana,oreg2010helical}. 
It was explicitly established that the combination of $s$-wave superconductivity induced in the SM by the SC, SU(2) spin symmetry breaking by the RSOC in the SM, and the breaking of time-reversal invariance by the Zeeman (or exchange) field in the SM gives rise to a topological superconductivity in the SM under well-defined conditions for the SM parameters~\cite{sau2010generic}. In particular, a 2D SM layer under such conditions supports localized MZMs inside ordinary vortex cores and the edge of the 2D SM layer represents a chiral Majorana wire with delocalized Majorana fermions running around the edge. It was quickly realized then that if the 2D SM is replaced by a semiconductor nanowire in the same setup of a SM-SC hybrid structure, the size quantization imposed by going from 2D to 1D converts the 1D nanowire into a helical Majorana wire with localized MZMs at the two ends of the wire; i.e., the delocalized chiral Majorana modes running around the edge of the 2D SM become end-localized bound MZMs in the 1D helical nanowire [here, {\it helical} refers to the fact that the low-energy states in the nanowire are spin-momentum locked, and therefore effectively spinless, see Fig.~\ref{fig:setup}(b)]. In such a SM nanowire situation, the localized MZMs arise naturally as bound quasiparticles at the wire ends with no need for inducing any vortex excitations. This system is reminiscent of the Kitaev chain idealization for topological superconductors carrying MZMs although there are significant differences between the two since the electrons in the nanowire are spinful, and the nanowire is equivalent to the Kitaev model only in the unphysical limit of a very high spin splitting~\cite{pan2023majorana}. Of course, the real significance is that the SM-SC nanowire hybrid structure is experimentally realizable in the laboratory in contrast to the Kitaev chain. Indeed, following these theoretical suggestions~\cite{sau2010generic,sau2010nonabelian,lutchyn2010majorana,alicea2010majorana,oreg2010helical}, there has been considerable experimental activity on SM-SC hybrid nanowires for the realization of non-Abelian MZMs over the last 15 years, with the experimental activity accelerating recently with Microsoft focused on this system as their (topological) quantum computing platform~\cite{microsoftquantum2023inasal,aghaee2025interferometric,aasen2025roadmap}.

We emphasize that RSOC plays the crucial role in taking a spinful $s$-wave SC and converting it into an effectively spinless $p$-wave SC in Majorana nanowire structures, but the physics is highly nontrivial. This can be seen by quoting from the original work of Kitaev~\cite{kitaev2001unpaired}, where it is explicitly stated that ``(p)hysical realization of a (topological) quantum wire is a difficult task because electron spectra are usually degenerate with respect to spin. Thus spin-orbit interaction does not help." The problem is the fermion doubling theorem which makes it impossible to realize topological SC using regular spinful $s$-wave SC. What the new ideas of 2009-2010 achieved is combining three separate physical mechanisms which are intrinsic to semiconductors with SIA: proximity $s$-wave SC, time reversal invariance breaking, and Rashba spin-orbit coupling. While RSOC separates the two spin bands in momenta (which by itself does not escape the fermion doubling problem), the spin splitting allows the avoidance of fermion doubling by tuning the spin gap and the SC gap. Figure~\ref{fig:setup} shows a schematic of the relevant physics and the key role of RSOC in creating topological SC in the SM-SC hybrid nanowires. We emphasize that not only is the Rashba effect essential in the emergence of topological superconductivity and MZMs in the engineered SM-SC hybrid structures, but the topological gap in the system is proportional to the strength of the Rashba coupling, and hence the fact that the Rashba coupling itself can be enhanced by engineering the actual geometry of the sample (e.g., by ensuring maximum spatial asymmetry in the nanowire) plays a key role in the physics---RSOC is essential, and it can be physically engineered to enhance the topological gap in order to make the MZMs more robust. This key role of the Rashba effect led to the theoretical predictions of using InAs or InSb as the SM material since these are the semiconductors known to have large Rashba coupling~\cite{sau2010generic,sau2010nonabelian,lutchyn2010majorana,alicea2010majorana,oreg2010helical}. Both InAs and InSb nanowires have been used in the experimental work on the subject with Microsoft focusing on InAs since it has better materials properties.

It is also worth mentioning that an alternative concrete proposal for the realization of 1D topological superconductivity in a discrete chain of proximitized quantum dots rather than a continuous nanowire was introduced in Ref.~\cite{sau2012realizing}. Recently, there has been impressive experimental progress towards verifying this idea in minimal two- and three-dot devices~\cite{dvir2023realization,Bordin2025,tenHaaf2024,tenHaaf2025}. Just like in the continuous nanowire case, the RSOC plays a crucial role in generating an effective $p$-wave superconducting pairing also in this quantum-dot based platform.

The fact that RSOC plays a role in SC properties has been known for a while. In particular, RSOC enables a strong enhancement of the so-called Chandrasekhar-Clogston limit in superconductors by increasing the critical magnetic field necessary for quenching spinful $s$-wave SC where, without any SOC, the critical spin splitting is basically the SC gap since any splitting above the gap prevents Cooper pairing of opposite spin electrons by definition. But the presence of RSOC, e.g., in noncentrosymmetric materials, mixes spin singlet and spin triplet SC, enhancing considerably the critical field, thus making the resultant SC more robust to spin splitting since the SC in the presence of RSOC is no longer just a singlet SC. This physics plays a crucial role in the SM-SC hybrid Majorana platforms as the TQPT here coincides precisely with the critical spin splitting for the $s$-wave SC without any RSOC---if there is no RSOC in the SM, all that would happen is that the proximity-induced SC from the superconductor would simply be quenched when the spin splitting equals the induced gap. In the presence of RSOC, however, the $s$-wave gap indeed vanishes at this point, but upon increasing the spin splitting beyond the TQPT, a new gap opens which is effectively a helical gap for an effective spinless $p$-wave topological SC. Although the RSOC plays no role in determining the TQPT, it plays a crucial role in determining the topological gap beyond the TQPT with the topological gap being proportional to the Rashba coupling. In fact, Rashba himself pointed out the key importance of RSOC in noncentrosymmetric SC with the resultant RSOC-induced spin degeneracy lifting and the mixing of singlet and triplet pairing, although no connection was made to any topological SC or MZMs in this work of Rashba~\cite{gor2001superconducting}.

Historically, the first proposals for combining the spin-orbit coupling with $s$-wave superconductivity in order to produce an effective spinless $p$-wave or $p_x + ip_y$ superconductivity were proposed independently in 2D topological insulators~\cite{fu2008superconducting} and in fermionic cold atom systems~\cite{zhang2008superfluid}. Although these 2D systems are in principle time reversal invariant, in contrast to the SM-SC hybrid structures with strong RSOC and induced spin-splitting where time reversal invariance is explicitly broken~\cite{sau2010generic}, both of these are platforms where the MZMs would have to be localized in some vortex cores. There has also been a separate proposal for Majorana modes in cold atom fermions in the presence of an applied magnetic field~\cite{sato2009nonabelian}. In all of these platforms, the topological superconducting phase is an effective spinless $p$-wave superconductor because of the SO coupling producing a single Fermi surface with a spin helix as shown in Fig.~\ref{fig:setup}.

We emphasize that this article is not an exhaustive review of the subject of MZMs, topological superconductivity, topological quantum computation, and/or semiconductor-superconductor Majorana nanowires although we touch upon these topics as necessary. Many such reviews already exist spanning the last 20 years~\cite{nayak2008nonabelian,wilczek2009majorana,leijnse2012introduction,alicea2012new,stanescu2013majorana,beenakker2013search,sarma2015majorana,elliott2015colloquium,sato2016majorana,aguado2017majorana,sato2017topological,oppen2017topological,lutchyn2018majorana,flensberg2021engineered,sau2021topological,laubscher2021majorana,marra2022majorana,dassarma2023search,yazdani2023hunting,amundsen2024colloquium,kouwenhoven2025perspective}. This article's focus is the role of the Rashba effect in creating topological superconductivity in artificially engineered structures where a regular metallic superconductor (e.g., Al) provides the superconductivity, but the main physics is in a material with strong Rashba coupling (a 1D nanowire or a 2D layer or a Josephson junction), and the focus here is on the role of the Rashba spin-orbit coupling in producing the topological superconductivity. We also mention in this context that the complementary topic of superconductivity and spin-orbit coupling in noncentrosymmetric (mostly bulk) materials, where the Rashba effect plays a central role by mixing singlet and triplet pairing (and thus enhancing the critical spin splitting associated with the Chandrasekhar-Clogston-Pauli limit) has also been well-reviewed in the literature~\cite{smidman2017superconductivity}.

The rest of this article is organized as follows. In Sec.~\ref{sec:2d}, we briefly review the historical developments leading to the first proposals for engineered topological superconductivity in 2D SM-SC hybrid structures and introduce the basic idea of how the combination of RSOC, Zeeman splitting, and superconducting proximity effect leads to an effective spinless $p$-wave superconductivity. In Sec.~\ref{sec:nanowire}, we introduce the simplified 1D version of this proposal, where the 2D semiconductor is replaced by a 1D Rashba nanowire, and discuss how information about the RSOC strength can be extracted from the curvature of the gap closing as a function of Zeeman field. Next, in Sec.~\ref{sec:hole_nanowire}, we show that not only electron-doped but also hole-doped SM nanowires could be used as a platform for MZMs due to the strong SOC of the SM valence band holes, focusing on gate-defined nanowires fabricated from 2D germanium (Ge) hole gases as a particularly promising example. We then move on to discuss planar Josephson junctions based on SM two-dimensional electron gases (2DEGs) with strong RSOC in Sec.~\ref{sec:JJ}, where we first briefly review the basic theory leading to the emergence of MZMs in these systems and then present some more recent results related to geometric effects (which, in particular, influence the RSOC strength) and the superconducting diode effect. Finally, in Sec.~\ref{sec:TI_JJ}, we comment on Josephson junctions based on topological insulator surface states as an alternative platform for MZMs. We conclude in Sec.~\ref{sec:conclusions} with a summary and a brief overview.

\section{2D structures}
\label{sec:2d}

The idea of non-Abelian statistics arising in superconductors originated in the work of Read and Green~\cite{read2000paired}, which showed that vortices in spinless chiral $p$-wave superconductors could support Majorana zero modes. It was also realized that for spinful superconductors, which are more realistic experimentally, one would actually need more complicated defects---so-called half-quantum vortices---to host MZMs. Soon afterwards, Ivanov showed explicitly that these half-quantum vortex modes indeed support non-Abelian statistics~\cite{ivanov2001non}. While interesting, this effort to demonstrate MZMs turned out to be a complicated search first to find candidate chiral $p$-wave superconductors such as Sr$_2$RuO$_4$~\cite{mackenzie2003superconductivity,Maeno2024} and then to find half-quantum vortices in these systems~\cite{jang2011observation}. 

A breakthrough approach for resolving this challenge was made by Gorkov and Rashba~\cite{gor2001superconducting}, who pointed out that two-dimensional superconductivity with inversion symmetry-breaking spin-orbit coupling similar to RSOC would generate a chiral $p$-wave component of superconductivity even in a conventional $s$-wave superconductor. This led to an active search for topological superconductivity in noncentrosymmetric superconductors~\cite{lu2008zero,sato2009topological,smidman2017superconductivity} in a variety of materials. The search for topological superconductivity in such intrinsic noncentrosymmetric superconductors is complicated and remains an ongoing effort.

The idea~\cite{gor2001superconducting} of using spin-orbit coupling to hybridize singlet and triplet superconductivity also percolated into the field of synthetic systems of ultra-cold atoms~\cite{zhang2008superfluid}, where it was hoped that the enhanced flexibility might resolve the constraints of having to discover new materials. Specifically, it was realized that adding RSOC in an ultra-cold atomic Fermi gas with $s$-wave Feshbach resonance leads to a chiral $p$-wave component in the interaction. In the presence of Zeeman splitting, such a $p$-wave interaction can emulate a spinless $p$-wave superconductor~\cite{zhang2008superfluid} along the lines proposed by Read and Green~\cite{read2000paired}. This was followed up by a calculation showing that a spin-orbit defect in an ultra-cold atomic gas can bind Majorana fermions~\cite{sato2009nonabelian}.

A key development in the context of solid state systems was the idea that the materials constraints of requiring superconductivity and Zeeman splitting could be relieved by considering proximity effects from both superconductors and magnetic insulators. Specifically, these proximity effects could be used to realize MZMs in defects in quantum spin Hall edges~\cite{fu2010electron} as well as on the surface of a three-dimensional topological insulator (TI)~\cite{fu2008superconducting}. The latter system, without the superconducting proximity effect, contains a surface Hamiltonian that is purely Rashba spin-orbit coupling without the conventional parabolic energy dispersion. In fact, the spin-momentum locking at the Fermi surface, which was crucial to creating chiral $p$-wave superconductivity in the proposal in ultra-cold atomic gases~\cite{zhang2008superfluid}, is a defining characteristic of the TI surface state.
This was used to show the existence of topological superconductivity on the surface of a TI~\cite{fu2008superconducting}.
However, this idea relies on the unique properties of the TI surface state, which is composed of a non-degenerate Fermi surface with spin-momentum locking without time-reversal symmetry breaking.

The necessity of a TI in generating a single Fermi surface can be circumvented if one were to break time-reversal invariance, for example, by proximity effect from a magnetic insulator (MI). This led to the idea that a broad class of semiconductors could be used to create a synthetic topological superconductor by proximitizing the semiconductor both with a magnetic insulator and a superconductor~\cite{sau2010generic}, which is shown schematically in Fig.~\ref{fig:setup}(a). The proximity effect from the magnetic insulator would produce a spin-polarized electron gas with a single Fermi surface needed for spinless $p$-wave superconductivity similar to the ultra-cold atomic gas proposal~\cite{zhang2008superfluid}. The superconductivity would now be proximity-induced, except that Pauli depairing would lead to an effectively non-superconducting state since the spin-singlet Cooper pairs in the superconductor are blockaded from entering the spin-polarized semiconductor. Introducing RSOC then creates a chiral $p$-wave type pairing amplitude proportional to the strength of the RSOC~\cite{sau2010nonabelian}. The RSOC is thus crucial here since a spin splitting by itself would not allow any SC proximity effect in the SM.  The key idea~\cite{sau2010generic} is the combined presence of spin splitting, SO coupling, and SC---all three ingredients must be present simultaneously.

\begin{figure*}
\centering
\includegraphics[width=0.95\textwidth]{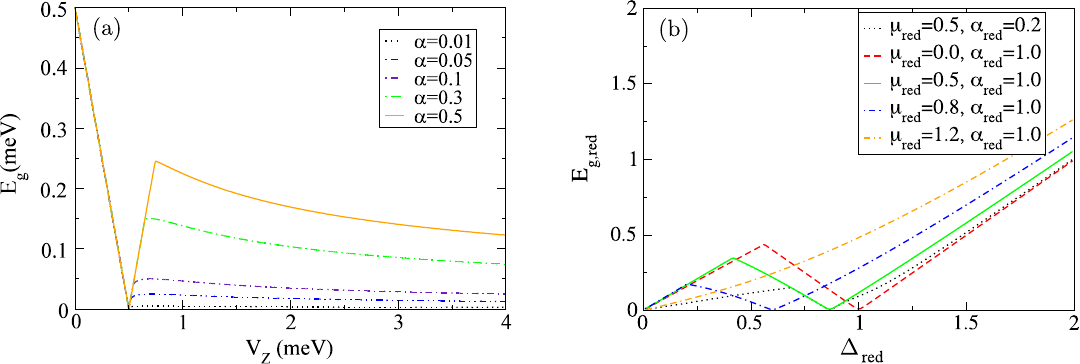}
\caption{\label{fig:gap} (a) Quasiparticle gap $E_g$ versus Zeeman splitting $V_Z$ for various strengths of the spin-orbit coupling $\alpha$, which is given in units of $0.3$~eV\,\AA\ here. The superconducting pairing potential and chemical potential are taken to be $\Delta=0.5$~meV and $\mu=0.0$~meV, respectively. The gap vanishes at a critical Zeeman splitting $V_{Z,c}=\sqrt{\Delta^2+\mu^2}$. For $V_Z<V_{Z,c}$, the superconducting gap is of conventional $s$-wave type. In the absence of RSOC, the superconducting gap is destroyed for $V_Z>V_{Z,c}$. In the presence of RSOC, on the other hand, a gap is opened up in this phase, leading to re-entrant superconductivity which is topological. (Adapted from Ref.~\cite{sau2010nonabelian}.) (b) Quasiparticle gap $E_g$ versus superconducting pairing potential $\Delta$ for various values of the chemical potential $\mu$. Here, $E_{g,\mathrm{red}}=E_g/V_Z$, $\Delta_{\mathrm{red}}=\Delta/V_Z$, $\mu_{\mathrm{red}}=\mu/V_Z$, and $\alpha_{\mathrm{red}}=\alpha/\sqrt{V_Z/2m^*}$. Again, the gap closes at the critical point, marking the phase transition between the topologically nontrivial (left) and the topologically trivial (right) phase. The maximum value of the gap in the topologically nontrivial phase, and the corresponding area in the phase diagram, decreases with increasing values of $\mu$. (Adapted from Ref.~\cite{sau2010generic}.)
}
\end{figure*}

A minimal Hamiltonian describing the SM-SC-MI heterostructure sketched in Fig.~\ref{fig:setup}(a) can be conveniently written in Bogoliubov-de Gennes (BdG) form as $H=\frac{1}{2}\int d^2\boldsymbol{r}\,\Psi^\dagger(\boldsymbol{r}) \mathcal{H}(\boldsymbol{r})\Psi(\boldsymbol{r})$ with the Nambu spinor $\Psi(\boldsymbol{r})=(\psi_\uparrow(\boldsymbol{r}),\psi_\downarrow(\boldsymbol{r}),\psi_\downarrow^\dagger(\boldsymbol{r}),-\psi_\uparrow^\dagger(\boldsymbol{r}))^T$ and 
\begin{align}
\mathcal{H}(\boldsymbol{r})&=\left[\left(-\frac{\nabla^2}{2m^*}-\mu\right)-i\alpha (\hat{z}\times\bm\sigma)\cdot\nabla\right]\tau_z+V_Z\sigma_z \nonumber\\
&\quad+[\Delta(\boldsymbol{r})\tau_++\mathrm{H.c.}].\label{eq:rashba_2deg}
\end{align}
In this equation, the first line denotes the SM Hamiltonian in the presence of RSOC and spin splitting without any superconductivity, whereas the second line is the SC proximity effect induced by the bulk superconductor in contact with the SM. Explicitly, $\psi_\sigma(\boldsymbol{r})$ is the fermion annihilation operator for the spin state $\sigma$, $m^*$ is the effective mass of the electrons, $\mu$ is the chemical potential, $\alpha$ is 
the strength of the RSOC, $V_Z$ is the Zeeman splitting, $\Delta(\boldsymbol{r})$ is the superconducting pairing potential, and $\sigma_{i}$ ($\tau_{i}$)  for $i\in\{x,y,z\}$ are Pauli matrices acting in spin (particle-hole) space. We have also defined $\tau_\pm=(\tau_x\pm i\tau_y)/2$ and set $\hbar=1$.
The quasiparticle excitation spectrum of $H$ can then be obtained by diagonalizing the BdG Hamiltonian $\mathcal{H}$. (The situation going beyond the mean field BdG theory has been investigated theoretically, and the physics remains the same~\cite{sau2011number,levin2020rigorous}.) The quasiparticle gap for the case of a uniform superconducting potential $\Delta(\boldsymbol{r})\equiv\Delta$ as a function of the parameters $V_Z$, $\Delta$, and $\alpha$ is shown in Fig.~\ref{fig:gap}. The quasiparticle gap as a function of the Zeeman potential $V_Z$ seen in Fig.~\ref{fig:gap}(a) shows a gap closing and reopening indicating a topological quantum phase transition at a critical Zeeman splitting $V_{Z,c}=\sqrt{\Delta^2+\mu^2}$~\cite{sau2010generic}. We note that the TQPT is unaffected by the RSOC. For $V_Z<V_{Z,c}$, the superconducting gap is of conventional $s$-wave type, while a topologically nontrivial gap reopens for $V_Z>V_{Z,c}$. This reopening of a nontrivial SC gap is made possible by the presence of RSOC---indeed, the conventional $s$-wave superconducting gap is completely destroyed as the Zeeman splitting exceeds the critical value $V_{Z,c}$, such that a reopening of the superconducting gap is only possible in the presence of a finite $p$-wave pairing amplitude induced by RSOC. Figure~\ref{fig:gap}(a) confirms that the size of the topological gap increases with increasing RSOC strength. The form of the critical point is further verified in Fig.~\ref{fig:gap}(b), which shows the quasiparticle gap as a function of the induced pairing potential $\Delta$ for different values of chemical potential $\mu$. Again, we see that the topological gap closes at the critical point, marking the phase transition between the topologically nontrivial and the topologically trivial phase. The maximum value of the gap in the topologically nontrivial phase, and the corresponding area in the phase diagram (not shown here), decrease with increasing values of $\mu$. Finally, we note that the topological superconducting phase can be confirmed explicitly by studying the excitation spectrum of a vortex in the superconducting pairing potential $\Delta(\boldsymbol{r})$, which can be shown to support an MZM in the vortex core only if the system is in the topological phase~\cite{sau2010generic}.

We conclude this section by pointing out that the RSOC also plays a crucial role in protecting the topological superconducting gap from disorder. Unlike conventional $s$-wave superconductors, time-reversal breaking chiral $p$-wave superconductors are susceptible to disorder when the mean-free path becomes shorter than the coherence length. Using an analysis of the superconductor with disorder~\cite{sau2012experimental} it was found that the critical strength of disorder that destroys superconductivity is proportional to the strength of Rashba spin-orbit splitting relative to the Zeeman splitting. We also mention that, apart from Rashba spin-orbit coupling, semiconductors can additionally support Dresselhaus spin-orbit coupling, which depends on the symmetry of the semiconductor quantum well. This degree of freedom has been proposed to be used to eliminate the requirement of a magnetic insulator. Specifically, it was shown that for the appropriate fine-tuned combination of Rashba and Dresselhaus spin-orbit coupling the topological superconducting phase can be realized with Zeeman splitting generated by an in-plane magnetic field~\cite{alicea2010majorana}.

\section{1D semiconductor- superconductor hybrid nanowires}
\label{sec:nanowire}

\subsection{Basic model}

The two-dimensional model introduced in the previous section has a simpler 1D realization based on a semiconductor-superconductor (SM-SC) hybrid nanowire~\cite{lutchyn2010majorana,oreg2010helical}, see Fig.~\ref{fig:setup}(b) for a schematic depiction of this setup. Via the essentially same mechanism as in the 2D case, the RSOC, in combination with Zeeman splitting and proximitiy-induced $s$-wave superconductivity, effectively creates a spinless $p$-wave component of the proximity-induced superconducting pairing in the SM nanowire if the chemical potential is tuned into the helical gap opened by the Zeeman field. This effectively turns the nanowire into an artificial 1D $p$-wave superconductor manifesting Majorana zero modes at the wire ends. Indeed, in the limit of infinite spin splitting, the nanowire becomes a spinless $p$-wave superconductor~\cite{sengupta2001midgap}, which, within a lattice approximation, is equivalent to the Kitaev chain model~\cite{kitaev2001unpaired,pan2023majorana}.

\begin{figure*}[ht]
    \centering
    \includegraphics[width=6.8in]{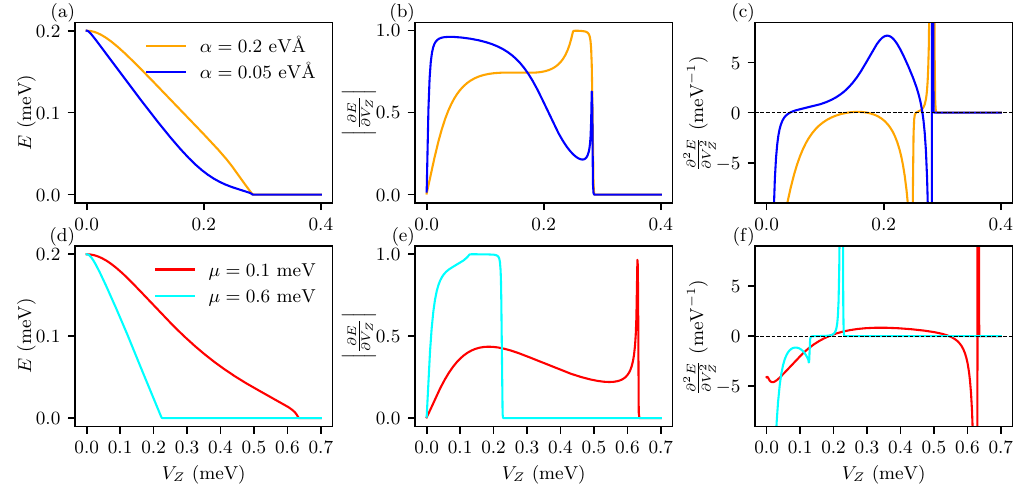}
    \caption{
    (a) Energy of the lowest-energy state, (b) magnitude of first derivative, (c) second derivative as a function of Zeeman field $V_Z$ for a weak RSOC $\alpha=0.05$ eV\AA{} (blue) and a strong RSOC $\alpha=0.2$ eV\AA{} (orange) at $\Delta=0.2~$meV, $\mu=0.2~$meV, and wire length $L=30~\mu$m. The orange line is mostly concave while the blue line is mostly convex.
    (d) Energy of the lowest-energy state, (e) magnitude of first derivative, (f) second derivative as a function of Zeeman field $V_Z$ for a small chemical potential $\mu=0.1$ meV (red) and a large chemical potential $\mu=0.6$ meV (cyan) at $\Delta=0.2$ meV, $\alpha=0.2$ eV\AA{}, and $L=30~\mu$m. (Adapted from Ref.~\cite{pan2019curvature}.)
    }
    \label{fig:E_vz}
\end{figure*}

The minimal 1D model describing the SM-SC hybrid nanowire is given by the BdG Hamiltonian [cf. Eq.~(\ref{eq:rashba_2deg}) above for the 2D case]
\begin{equation}
H=\frac{1}{2}\int dx\, \Psi^\dagger(x) \left( H_{\text{SM}}+H_{\text{Z}}+H_{\text{SC}} \right) \Psi(x),\label{eq:rashba_nanowire}
\end{equation}
where $H_{\text{SM}}$, $H_{\text{Z}}$, and $H_{\text{SC}}$ describe the semiconductor with RSOC, the Zeeman field, and the proximity-induced superconductivity, respectively. 
With the choice of Nambu spinor $\Psi(x)=(\psi_\uparrow(x),\psi_\downarrow(x),\psi_\downarrow^\dagger(x),-\psi_\uparrow^\dagger(x))^T$, the single-particle (first-quantized) Hamiltonian for the three terms can be written as
\begin{equation}\label{eq:H}
\begin{split}
H_{\text{SM}}&=\left( -\frac{\partial_x^2}{2m^*}-i\alpha \partial_x\sigma_y-\mu \right)\tau_z,\\
H_{\text{Z}}&=V_Z \sigma_x,\\
H_{\text{SC}}&=\Delta \tau_x,
\end{split}
\end{equation}
where all symbols have the same meaning as in Eq.~(\ref{eq:rashba_2deg}). The Hamiltonian terms satisfy the BdG particle-hole symmetry represented by $\mathcal{P}=\tau_y\sigma_y \mathcal{K}$ for our choice of Nambu basis, where $\mathcal{K}$ is the complex conjugation. In terms of materials, typical choices for the SM nanowire are InAs or InSb due to their strong RSOC strength and large $g$-factors, whereas the SC is taken to be Al or Nb. (The large $g$ factor is a practical necessity so that a weak magnetic field can be applied---otherwise, the applied magnetic field will quench the superconductivity in the parent superconductor and suppress all proximity effect before the TQPT is reached.) For InSb/Al hybrids, typical material parameters are $m^*=0.015\,m_e$ for the effective mass (where $m_e$ is the electron rest mass), $\alpha=0.1-1$ eV\AA{} for the RSOC strength, $\mu=0-1$ meV for the chemical potential, and $\Delta=0.2$ meV for the proximity-induced superconducting gap~\cite{lutchyn2018majorana}. We note that an important difference compared to the 2D setup discussed in the previous section is that the required spin splitting in the 1D system can be created by a Zeeman field oriented along the nanowire axis, which can be induced by a magnetic field. This is not possible in the 2D case, where the magnetic field would have to be oriented in the direction perpendicular to the plane, which would almost immediately destroy superconductivity by orbital effects due to the large cross section of the superconductor penetrated by the magnetic field.

A convenient method to study the Hamiltonian given in Eq.~(\ref{eq:rashba_nanowire}) for a wire of finite length is to discretize the Hamiltonian on a grid in real space using the finite-difference method. The low-energy spectrum of the resulting effective tight-binding Hamiltonian can then readily be obtained via exact numerical diagonalization. This strategy can, for example, be used to verify that the closing and reopening of the bulk gap at the TQPT, which is occurring at $V_{Z,c}=\sqrt{\mu^2+\Delta^2}$ just like in the 2D case, is accompanied by the emergence of zero-energy MZMs exponentially localized at the wire ends for sufficiently long wires~\cite{lutchyn2010majorana,sau2010generic,sau2010nonabelian,oreg2010helical}. As another example related directly to the important role of the RSOC in the SM-SC hybrid Majorana nanowire, we will use the numerical tight-binding approach to estimate the strength of the RSOC from the curvature of the bulk gap closing in the next subsection.

We note that Eq.~\eqref{eq:H} is based on several key approximations: First, the proximity-induced superconductivity is modeled as a constant pairing amplitude $\Delta$ in the semiconductor, which is a convenient approximation and generally valid in the limit of weak coupling between the SM and the SC. In the intermediate-to-strong coupling limit, one can explicitly model the tunneling of electrons across the SM-SC interface and then integrate out the SC degrees of freedom in order to obtain an effective energy-dependent self-energy in the Green's function of the total Hamiltonian~\cite{stanescu2010proximity,stanescu2013majorana,Tewari2011}, which can then be solved self-consistently. However, this energy-dependent self-energy term only quantitatively renormalizes the proximitized SC gap in the semiconductor and does not qualitatively change the low-energy physics of interest here. Additionally, the collapse of the parent superconducting gap $\Delta_0$ due to the magnetic field can be taken into account in this approach by introducing a phenomenological $V_Z$-dependent reduction $\Delta_0(V_Z)=\Delta_0(V_Z=0)\sqrt{1-(V_Z/V_Z^*)^2}$, where $V_Z^*$ is the critical Zeeman field at which the parent superconducting gap is found to collapse in experiments. However, this effect is negligible as long as we consider Zeeman fields far below $V_Z^*$. Second, we have focused on the ideal pristine case ignoring the presence of unintentional random disorder in the nanowire since, at the theoretical level, the physics of the RSOC---which is the focus of this article---is already captured in the clean limit. However, the disorder contribution is important in actual experiments and has been extensively discussed in the literature~\cite{brouwer2011topological,brouwer2011probability,liu2012zerobias,bagrets2012class,akhmerov2011quantized,sau2013density,takei2013soft,adagideli2014effects,pekerten2017disorderinduced,brzezicki2017driving, barmanray2021symmetrybreaking,pan2020physical,ahn2021estimating,woods2021chargeimpurity,dassarma2021disorderinduced,dassarma2023density,dassarma2023spectral,dassarma2023search,pan2024disordered,taylor2024machine}. Third, we work in the single-band approximation, where the occupancy of the wire is low without strong inhomogeneities. For a discussion of multiband models, where the basic MZM physics remains the same, we refer the reader to Refs.~\cite{lutchyn2011search,stanescu2011majorana}. Finally, we mention that, while the standard Majorana nanowire system discussed here consists of a semiconductor nanowire that is \emph{partially} covered by a superconductor, so-called full-shell nanowires, where the nanowire is fully enclosed by a superconducting shell, have been considered as a potential platform for MZMs as well~\cite{Vaitiekenas2020,Penaranda2020}.

\subsection{Estimating the RSOC strength from the gap closing curvature}

The key role of the RSOC in the SM-SC nanowire is that it breaks spin SU(2) symmetry and creates an effective spinless $p$-wave component of the superconducting pairing whose strength is proportional to the RSOC strength $\alpha$ (in spite of the parent superconductor, e.g., Al, being strictly $s$-wave)~\cite{lutchyn2010majorana,pan2023majorana}.
Therefore, it is favorable to use semiconductor materials with a large predicted RSOC strength such as InSb ($\alpha\sim0.2-1$ eV\AA{}) or InAs ($\alpha\sim0.2-0.8$ eV\AA{})~\cite{lutchyn2018majorana}. 
However, the actual RSOC strength in the experimental hybrid nanowire structures is quantitatively unknown since it cannot be measured directly (the numbers reported in the literature are usually for bulk systems). With this motivation, we show that information about the strength of the RSOC can be obtained by tracking the curvature of the gap closing feature in the lowest-energy state of the SM-SC hybrid system. Intuitively, this is based on the idea that, for ideal wires, a larger RSOC strength will lead to more concavity in the gap-closing feature because the RSOC-induced triplet component of the SC pairing is robust in the presence of a magnetic field. 
Quantitatively, we can track the curvature via the second derivative of the lowest-energy state $E(V_Z)$ with respect to the Zeeman field $V_Z$ in the trivial phase; explicitly, $\frac{\partial^2E}{\partial V_Z^2}>0$ ($\frac{\partial^2E}{\partial V_Z^2}<0$) indicates a convex (concave) gap closing feature.

At zero RSOC, we expect the lowest energy to simply decrease linearly with the Zeeman field $V_Z$ until the gap closes at $V_Z=\Delta$, i.e., $E(V_Z)=\left( \Delta-V_Z \right) \theta\left(  \Delta-V_Z\right)$, where $\theta(x)$ is the Heaviside step function. This is the standard Pauli spin blockade of pairing in an $s$-wave singlet SC, where superconductivity must vanish when the spin splitting equals the SC gap in the absence of any SOC. Hence the second derivative is zero in the absence of RSOC except at $V_Z=\Delta$, i.e., ${\frac{\partial^2E}{\partial V_Z^2}=\delta(\Delta-V_Z)}$.
At a relatively weak RSOC, e.g., $\alpha=0.05$ eV\AA{}, see the blue lines in Figs.~\ref{fig:E_vz}(a-c), the gap closing feature is preponderantly convex, i.e., $\frac{\partial^2E}{\partial V_Z^2}>0$ for most of the values away from the TQPT and zero Zeeman field. As the strength of the RSOC increases further, e.g., $\alpha=0.2$ eV\AA{}, see the orange lines in Figs.~\ref{fig:E_vz}(a-c), the gap closing feature becomes completely concave, i.e., $\frac{\partial^2E}{\partial V_Z^2}<0$ for all finite Zeeman fields except for a tiny region near the TQPT due to finite size effects.

\begin{figure}[tb]
    \centering
    \includegraphics[width=3.4in]{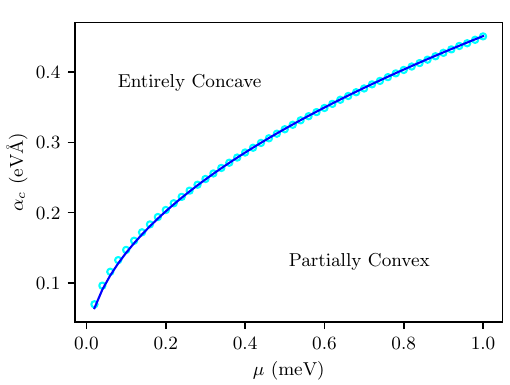}
    \caption{Dependence of the critical RSOC strength $\alpha_c$ on the chemical potential $\mu$. For $\alpha>\alpha_c$, the gap-closing feature is completely concave, while it is partially convex for $\alpha<\alpha_c$. Cyan points correspond to numerical data and the blue line corresponds to the analytical result given in Eq.~\eqref{eq:alpha_c}. In the numerical simulation, we have set $\Delta=0.2$ meV. (Adapted from Ref.~\cite{pan2019curvature}.)
    }
    \label{fig:alpha_c}
\end{figure}

The critical RSOC strength $\alpha_c$ beyond which the gap closing feature becomes entirely concave also depends on the chemical potential $\mu$ as shown in Fig.~\ref{fig:E_vz}(d-f), where we fix the RSOC strength $\alpha=0.2$ eV\AA{} and vary the chemical potential $\mu$.
We find that larger chemical potentials require a larger RSOC strength to reach the critical point. 
This dependence of the critical RSOC strength $\alpha_c$ on the chemical potential $\mu$ can be analytically derived by considering the inflection point of the lowest-energy state $E(V_Z)$ being zero, and we find that~\cite{pan2019curvature}
\begin{equation}\label{eq:alpha_c}
    \alpha_c = \beta \sqrt{\frac{h^2\mu}{m^*}},
\end{equation}  
where $\beta$ is a dimensionless constant.
We numerically compute the critical RSOC strength $\alpha_c$ as a function of the chemical potential $\mu$ at fixed $\Delta=0.2$ meV, see the cyan dots in Fig.~\ref{fig:alpha_c}, and find excellent agreement with the analytical result given by Eq.~\eqref{eq:alpha_c} with $\beta \approx 0.10117$ as shown by the blue line.
This monotonic dependence of the critical RSOC strength $\alpha_c$ on the chemical potential $\mu$ provides an explanation for the gap closing feature in the lowest-energy state in Fig.~\ref{fig:E_vz}. Namely, as the RSOC strength $\alpha$ increases, the gap closing feature in the lowest-energy state changes from partially convex to completely concave, and a larger critical RSOC strength $\alpha_c$ is required to reach complete concavity as the chemical potential $\mu$ increases. This again demonstrates the crucial role of RSOC in Majorana nanowires, where the RSOC strength is invariably connected with the spin splitting, the SC gap, and the chemical potential.

\section{Hole nanowires}
\label{sec:hole_nanowire}

\begin{figure*}[tb]
    \centering
    \includegraphics[width=0.9\textwidth]{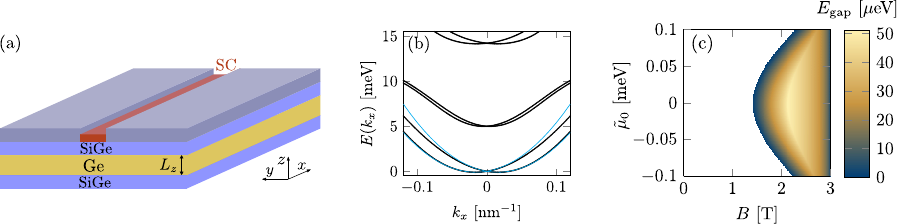}
    \caption{(a) A quasi-1D Ge hole nanowire defined by electrostatic gates (gray) placed on a Ge quantum well of thickness $L_z$ (yellow) sandwiched between two layers of SiGe (blue). If the nanowire is additionally proximitized by a superconductor (red) and a magnetic field is applied along the wire axis, MZMs can emerge at the wire ends. (b) Low-energy spectrum of the bare Ge hole nanowire at $B=0$ (note that a global minus sign is omitted from the hole spectrum throughout this work). The spectrum of the full (effective) model is shown in black (blue). (c) Bulk topological phase diagram for a Ge/Al hybrid nanowire. The white (colored) regions correspond to the trivial (topological) phase.  In (b) and (c), the parameters for the Ge hole nanowire are $L_y=15$~nm, $L_z=22$~nm, $\mathcal{E}=0.5$~V$/\mu$m, and $E_s=0.01$~eV. In (c), $\Delta(0)=80~\mu$eV and $B_c=3$~T. The chemical potential $\tilde\mu_0$ is measured from the spin-orbit point of the lowest confinement-induced subband. (Adapted from Ref.~\cite{laubscher2024germaniumbased}.)
    }
    \label{fig:ge_nanowire}
\end{figure*}

Electron-doped nanowires fabricated from InAs or InSb, as described in the previous section, are currently the most mature platform for the experimental search for MZMs. Alternatively, however, MZMs could in principle also be realized in hole-doped group IV semiconductors such as Ge~\cite{maier2014majorana,laubscher2024majorana,laubscher2024germaniumbased}. Indeed, narrow gate-defined wires fabricated from Ge 2D hole gases (2DHGs) could be a particularly promising Majorana platform due to the extraordinary (essentially disorder-free) quality of the existing Ge 2DHGs and their experimental compatibility with superconductors~\cite{scappucci2020germanium}.

The spin-orbit coupling physics is, however, more complex in hole systems. In hole systems, the uppermost valence band is described by wave functions originating from atomic $p$ orbitals with orbital angular momentum $l = 1$. Together with the usual spin $1/2$ this results in an effective spin $3/2$ degree of freedom of the valence band holes, which makes hole systems fundamentally different from electron systems. Nevertheless, in certain regimes, the band structure of Ge hole nanowires has similarities to the band structure of standard electron nanowires. In particular, structural inversion asymmetry controlled by, e.g., an applied electric field can lead to a sizable and tunable RSOC for Ge holes. Just like in the case of electron nanowires, this RSOC should in principle enable a topological superconducting phase with MZMs in Ge-based SM-SC hybrid nanowires.

To illustrate this, we consider a gate-defined Ge hole nanowire in a Ge/SiGe quantum well proximitized by a superconductor (such as, e.g., Al), see Fig.~\ref{fig:ge_nanowire}(a) for a schematic. We model the Ge hole nanowire by the Hamiltonian $H_0=\int d\bm{r}\,\psi^\dagger(\bm{r})\mathcal{H}_0(\bm{r})\psi(\bm{r})$ with $\psi=(\psi_{3/2}$,$\psi_{1/2},\psi_{-1/2},\psi_{-3/2})^T$ and
\begin{align}
\mathcal{H}_0&=\frac{\hbar^2}{m}\left[
\left(\gamma_1 + \frac{5 \gamma_s}{2}\right)\frac{\bm{k}^{2}}{2}
- \gamma_s \left( \bm{k} \cdot \bm{J} \right)^2 
\right]-\mu_0\nonumber\\&\quad+V(y,z)-E_s J_z^2-e \mathcal{E} z+H_\mathbf{B},\label{eq:normalH}
\end{align}
where the first line is the isotropic Luttinger-Kohn (LK) Hamiltonian describing the topmost valence band holes of 3D bulk Ge~\cite{luttinger1956quantum,winkler2003spin}. Here, $m$ denotes the bare electron mass, $\gamma_1=13.35$, $\gamma_2=4.25$, and $\gamma_3=5.69$ are the Luttinger parameters for Ge, $\gamma_s=(\gamma_2+\gamma_3)/2$, $\bm{k}=(k_x,k_y,k_z)$ is the vector of momentum, $\bm{J}=(J_x,J_y,J_z)$ is the vector of spin-$3/2$ operators, and $\mu_0$ is the chemical potential. In the second line of Eq.~(\ref{eq:normalH}), the confinement potential $V(y,z)$ describes (i) the confinement along the $z$ direction arising due to the SiGe barrier defining the quantum well and (ii) the gate-induced confinement along the $y$ direction that defines the quasi-1D nanowire. We model this confinement potential as an infinite square well in both the $y$ and $z$ direction, $V(y,z)= 0$ for $|y|<L_y/2, \ |z|<L_z/2$ and $V(y,z)=\infty$ otherwise, where $L_y$ is the width of the 1D channel and $L_z$ is the thickness of the quantum well. We note that using, e.g., a parabolic potential to model the gate-induced confinement along the $y$ direction leads only to quantitative, but not qualitative, changes to the results presented in the following~\cite{laubscher2024majorana}. Furthermore, $E_s>0$ is the strain energy arising due to the lattice mismatch between the Ge and the SiGe barrier~\cite{bir1974symmetry}. Finally, two crucial ingredients for the realization of MZMs are, as discussed in the previous sections, spin-orbit coupling and time-reversal symmetry breaking. The former is induced by an out-of-plane electric field $\mathcal{E}$, causing SIA (and hence, RSOC), while the latter results from a magnetic field $\mathbf{B}=(B,0,0)$ applied along the wire, which can be described by a term $H_\mathbf{B}=\mathcal{H}_Z+\mathcal{H}_\mathrm{orb}$. Here, the Zeeman term $\mathcal{H}_Z$ takes the form $\mathcal{H}_Z=2\kappa\mu_B B J_x$~\cite{luttinger1956quantum,winkler2003spin} with $\mu_B$ the Bohr magneton and $\kappa\approx 3.41$ for Ge~\cite{lawaetz1971valence}, while $\mathcal{H}_\mathrm{orb}$ describes an extra contribution to the bulk LK Hamiltonian due to orbital effects (see e.g., Ref.~\cite{adelsberger2022enhanced} for its explicit form).

Finally, we incorporate a proximity-induced superconducting pairing term into our model. Since the precise microscopic description of the proximity-induced superconducting pairing in Ge/SC hybrid structures is likely to depend on the microscopic details of the setup, we work with a phenomenological pairing term of the form
\begin{equation}
H_{sc}=\int d\bm{r} \sum_{s=\frac{1}{2},\frac{3}{2}}\Delta_s\,\psi_s^\dagger(\bm{r}) \psi_{-s}^\dagger(\bm{r})+\mathrm{H.c.},\label{eq:sc_pairing}
\end{equation}
where $\Delta_{1/2}$ is the superconducting pairing amplitude for holes with spin projection $\pm 1/2$ along the $z$ direction (light holes, LHs) and $\Delta_{3/2}$ is the pairing amplitude for holes with spin projection $\pm 3/2$ along the $z$ direction (heavy holes, HHs). To keep the number of unknown parameters to a minimum, we assume that the LH and HH pairing amplitudes are equal in magnitude but of opposite sign, i.e., $\Delta_{3/2}=-\Delta_{1/2}\equiv\Delta$. Furthermore, we model the suppression of the proximity-induced superconducting gap due to the applied magnetic field as
%
%\begin{equation}
$\Delta(B)=\Delta(0)\sqrt{1-\left(B/B_c\right)^2}\,\Theta(B_c-|B|)$,
%\end{equation}
%
where $\Delta(0)$ is the proximity-induced superconducting gap at zero magnetic field, $B_c$ is the critical magnetic field of the superconductor, and $\Theta$ is the Heaviside step function. 

\begin{figure*}[tb]
    \centering
    \includegraphics[width=0.95\textwidth]{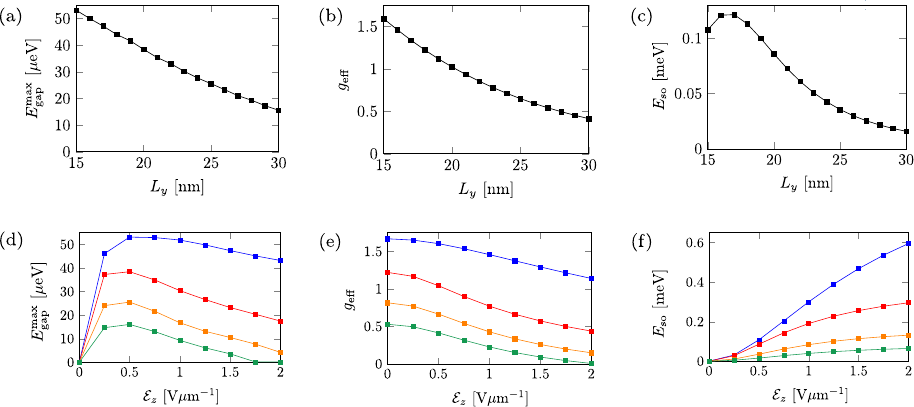}
    \caption{(a-c) Maximal bulk gap in the topological phase, effective $g$ factor at $B=2$~T, and effective spin-orbit energy at zero magnetic field in dependence on the channel width. (d-f) Maximal bulk gap in the topological phase, effective $g$ factor at $B=2$~T, and effective spin-orbit energy at zero magnetic field in dependence on the electric field for different channel widths (blue: $L_y=15$~nm, red: $L_y=20$~nm, orange: $L_y=25$~nm, green: $L_y=30$~nm). We fix $L_z=22$~nm, $\Delta=0.1$~meV, and $E_s=10$~meV for all panels. In panels (a-c), we have further used $\mathcal{E}_z=0.5$~V$\mu$m$^{-1}$. The solid lines are a guide to the eye only. (Adapted from Ref.~\cite{laubscher2024majorana}.)
    }
    \label{fig:ge_nanowire_params}
\end{figure*}

We can now study the full Hamiltonian $H=H_0+H_{sc}$ numerically. A convenient way to do this is by projecting the Hamiltonian onto a set of basis functions that solve the confinement problem, see e.g., Ref.~\cite{laubscher2024majorana} for details. In Fig.~\ref{fig:ge_nanowire}(b), we show the band structure for a bare (unproximitized) Ge hole nanowire that is assumed to be infinite along the $x$ direction such that $k_x$ is a good quantum number. We see that the spectrum of the lowest two subbands visually resembles the spectrum of a conventional Rashba nanowire up to a global minus sign, which we omit here. In Fig.~\ref{fig:ge_nanowire}(c), we show the corresponding bulk topological phase diagram obtained by numerically evaluating the Pfaffian invariant $\mathrm{Pf}(H)$ as well as the size of the topological gap. Here, the chemical potential is chosen such that only the lowest confinement-induced hole subband is occupied. The gap closing line marking the transition between the trivial phase (white) and the topological phase (colored) closely resembles the parabolic form known from electron nanowires (see Sec.~\ref{sec:nanowire}), a connection that we will elaborate more on below.

The physical origin of the hole SOC seen in Fig.~\ref{fig:ge_nanowire}(b) has been studied in detail in the context of hole spin qubits. This type of hole SOC is typically referred to as \emph{direct} RSOC~\cite{kloeffel2011strong,kloeffel2018direct} and can become very strong in hole systems with two axes of strong confinement such as nanowires and elongated quantum dots. Here, `direct' refers to the fact that the SOC is not suppressed by the fundamental band gap of the semiconductor, but arises from the four-band LK Hamiltonian by itself in the presence of an electric field. A theoretical analysis shows that the direct RSOC crucially depends on the HH-LH mixing induced by the strong confinement along two coordinate axes. Consequently, the direct RSOC is negligible in two-dimensional systems with only one axis of strong confinement, where the topmost valence band has almost exclusively HH character. Indeed, in a 2D hole gas, the leading spin-orbit term is typically not linear but cubic in momentum~\cite{scappucci2020germanium}. Similarly, compressive strain is known to increase the HH-LH splitting and is therefore generally detrimental to the direct SOC.

The similarities with the conventional electron Rashba nanowire become particularly transparent if we write down an effective model describing the two lowest-energy bands of the Ge hole nanowire, which can be labeled by a pseudospin index $\sigma\in\{\Uparrow,\Downarrow\}$. Up to a global minus sign we find that the low-energy band structure of the normal-state Hamiltonian $H_0$ is well described by a two-band model of the form $H_0^\mathrm{eff}=\int dx\,\psi^\dagger(x)\mathcal{H}_0^\mathrm{eff}\psi(x)$ with $\psi=(\psi_\Uparrow,\psi_\Downarrow)^T$ and~\cite{adelsberger2022enhanced}
\begin{equation}
\mathcal{H}_0^\mathrm{eff}=-\frac{\hbar^2 \partial_x^2}{2m^*}-\mu+\left(V_Z-\frac{\hbar^2\partial_x^2}{2m^*_s}\right)\sigma_x-i\alpha\partial_x\sigma_y,\label{eq:eff2band}
\end{equation}
where $m^*$ is the effective mass, $\alpha$ is the effective spin-orbit coupling strength, $m^*_s$ is an effective spin-dependent mass, and the Pauli matrices $\sigma_i$ with $i\in\{x,y,z\}$ act in pseudospin space. Furthermore, we have defined the effective Zeeman splitting $V_Z=g_\mathrm{eff}\mu_BB/2$, where $g_\mathrm{eff}$ is the effective $g$ factor. However, we stress again that, in contrast to spin-$1/2$ electrons in conventional Rashba nanowires, the low-energy holes considered here are a mix of HHs and LHs, with the degree of HH-LH mixing depending sensitively on the details of the system (e.g., wire geometry, strain, electromagnetic fields).

Since the effective parameters entering Eq.~(\ref{eq:eff2band}) are strongly geometry-dependent, so is the topological phase. As an example, in Fig.~\ref{fig:ge_nanowire_params}(a), we plot the maximal topological gap in dependence on the channel width $L_y$ and find that it decreases monotonically with increasing channel width. Furthermore, narrow channels also exhibit a significantly larger topological phase space (not shown here) than wide channels, where the topological phase can only be achieved at high magnetic fields close to the critical field of the superconductor. This can be understood as follows: First, as the width of the channel increases, the effective $g$ factor $g_\mathrm{eff}$ [see Fig.~\ref{fig:ge_nanowire_params}(b)] decreases significantly due to the decreasing HH-LH mixing. Indeed, it is well known that the effective in-plane $g$ factor of Ge holes becomes very small as one moves towards the 2D limit where the lowest subband has predominantly HH character~\cite{scappucci2020germanium}. Second, the spin-orbit energy $E_{so}=m^*\alpha^2/2\hbar^2$ [see Fig.~\ref{fig:ge_nanowire_params}(c)] reaches a maximum at a relatively small width ($L_y\approx 16$-$17$~nm) and decreases significantly as the width increases further. This is because, as one moves away from the strictly 1D limit with two axes of strongest confinement, the SOC becomes suppressed due to the decreasing HH-LH mixing~\cite{kloeffel2011strong,kloeffel2018direct}. In Figs.~\ref{fig:ge_nanowire_params}(d-f), we additionally show the maximal topological gap, the effective $g$-factor, and the spin-orbit energy as a function of the electric field, which can be used as a tuning knob to maximize the topological gap for a given wire geometry. We also note that the spin-dependent mass $m_s^*$ is very small in the parameter regime we are interested in, such that we neglect it.

The effective low-energy Hamiltonian describing the proximitized Ge hole nanowire in the single-subband regime then takes the exact same form as Eq.~(\ref{eq:H}). Therefore, many results from electron nanowires can be expected to carry over to hole nanowires, albeit with quantitative differences due to differences in materials parameters and a significant geometry-dependence of key quantities such as the effective $g$-factor and the hole SOC strength. In general, the main disadvantage of the Ge-based Majorana platform is the overall small $g$ factor, which limits both the topological phase space and the maximal topological gaps that can be achieved. This situation could be improved by careful band structure engineering and optimization of the device geometry, which might make it possible to increase the effective $g$ factor beyond the values considered here. Additionally, superconductors with a larger bulk gap than Al could be used (e.g., Nb), and the geometry of the superconducting thin film could be optimized to increase the critical field for a given superconductor. On the other hand, the main advantage of the Ge-based platform is the strong hole SOC and the extremely high quality (low disorder) of the existing Ge 2DHGs.

%%%%%%%%%%%%%%%%%%%%%%%%%%%%%%%%%%%%%%%%%%%%%%%%%%%%%%%%%%%%%%%%%%%%%%%%%%%%%%%%%%
%%%%%%%%%%%%%%%%%%%%%%%%%%%%%%%%%%%%%%%%%%%%%%%%%%%%%%%%%%%%%%%%%%%%%%%%%%%%%%%%%%

\section{Semiconductor-based planar Josephson junctions}           \label{sec:JJ}

One of the disadvantages of the nanowire platforms discussed in Secs.~\ref{sec:nanowire} and \ref{sec:hole_nanowire} is that relatively large Zeeman splittings exceeding the critical field $V_{Z,c}$ are needed to drive the system into the topological phase, requiring sizable magnetic fields that are detrimental to superconductivity. As a possible alternative to the standard nanowire setup, it was therefore suggested that a SM-SC hybrid Majorana platform could also be obtained by placing two extended $s$-wave superconducting leads on a semiconductor 2DEG with strong RSOC such that a planar Josephson junction (JJ) is formed~\cite{Hell2017,Pientka2017,Schiela2024}, see Fig.~\ref{Fig_V1}. Here, the narrow normal region between the two superconducting leads essentially plays the role of the `nanowire', which inherits superconducting correlations from the two leads. In the presence of a magnetic field, the JJ can then enter a topological superconducting phase manifesting MZMs at the two ends of the quasi-1D junction region. Furthermore, the superconducting phase difference constitutes an additional tuning knob to break time-reversal symmetry in the JJ platform, which has been theoretically predicted~\cite{Hell2017,Pientka2017} to allow for the emergence of topological superconductivity and MZMs at reduced magnetic fields compared to the standard nanowire setup. The crucial role of RSOC, however, remains similar to the nanowire situations discussed in the earlier sections.

In the following, we briefly review the main advances in the development of semiconductor-based planar JJs into a flexible platform for topological superconductivity and Majorana physics, starting with a discussion of the main idea and some basic models that describe the low-energy physics of the JJ heterostructure in Sec.~\ref{S5_1}. In Sec.~\ref{S5_2}, we then mention a few effects that emphasize the role of the ``effective geometry'' of the 2D structure, which incorporates constraints imposed by the patterning of the superconductors and the applied gate potentials. We conclude in Sec.~\ref{S5_4} with a discussion of the superconducting diode effect---a phenomenon in which spin-orbit coupling plays a key role and which could be used as a tool for estimating the SOC strength in hybrid structures. We emphasize that the planar JJ platform using a 2DEG is nowhere near as developed experimentally as the Majorana nanowire platform, and there are very few reported experimental measurements in this system. It is therefore possible that there are unknown problems in this platform which have not been appreciated yet.  Most of the work in this platform so far has been theoretical.

\subsection{Physical picture and basic modeling} \label{S5_1}

\begin{figure}[t]
\centering
\includegraphics[width=0.48\textwidth]{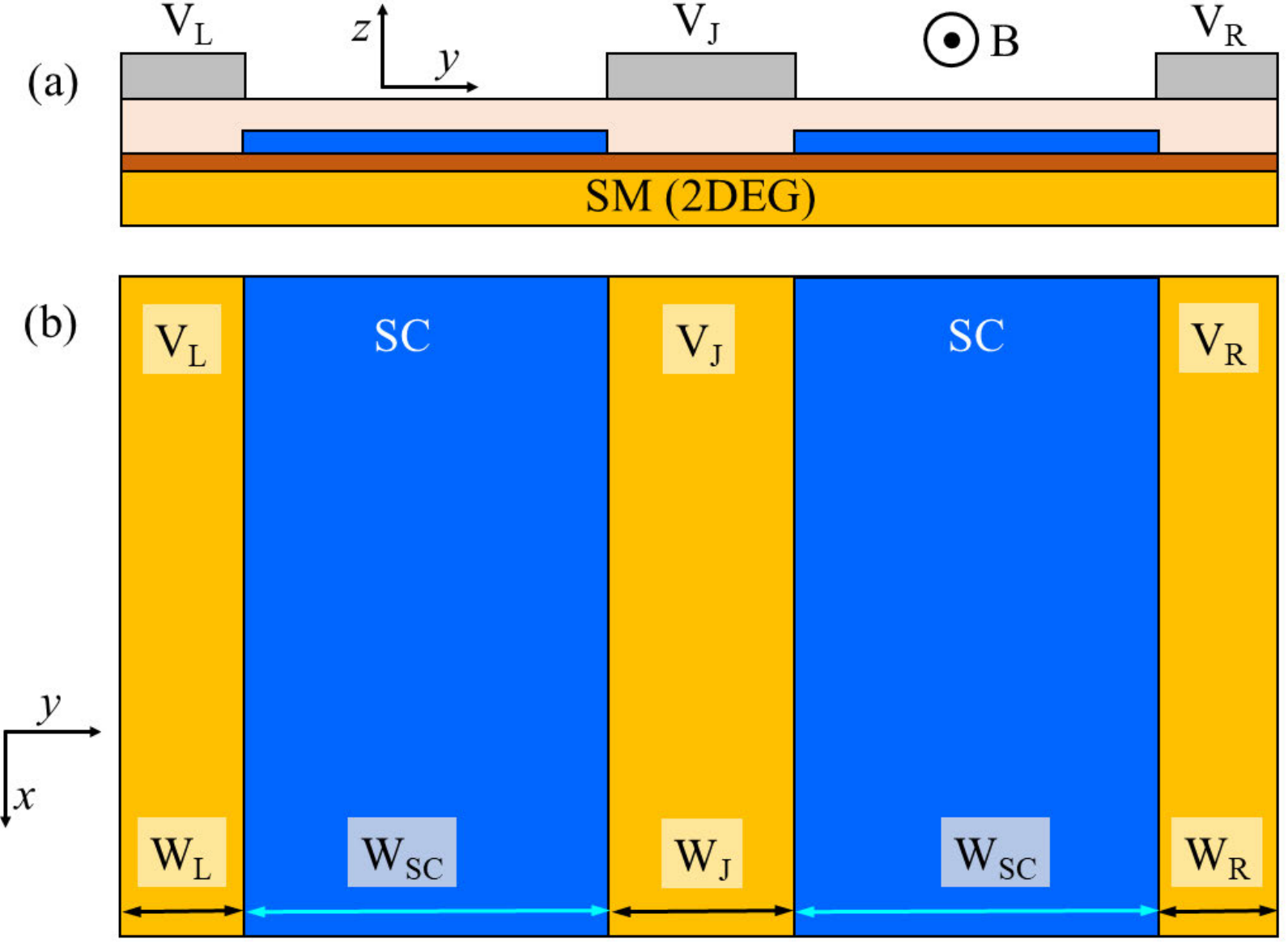}
\caption{Schematic representation of the planar Josephson junction device: (a) lateral view; (b) top view of the SM and SC components. A 2DEG with strong SOC hosted by a SM quantum well (orange) is proximity coupled to two thin SC films (blue) of width $W_{SC}$ and relative phase difference $\phi$, forming a Josephson junction of width $W_{J}$ and length $L$. A magnetic field $B$ is applied in the $x-y$ plane (typically parallel to the junction, i.e., in the $x$ direction).  Unproximitized SM regions (of widths $W_L$ and $W_R$) may be present outside the SC films. The electrostatic potential in the regions not covered by the SCs is controlled by top gates (gray), with $V_J$ representing the potential in the junction region.}
\label{Fig_V1}
\end{figure}

We consider a semiconductor-superconductor (SM-SC) hybrid system consisting of a planar semiconductor heterostructure that hosts a two-dimensional electron gas confined within a quantum well and proximity coupled to two superconducting films (see Fig.~\ref{Fig_V1}). The electrostatic potential within the regions not covered by the SC is controlled by top gates and a magnetic field is applied in-plane, typically parallel to the junction. 
The generic Hamiltonian describing the hybrid system has the form
\begin{equation}
H = H_{SM} + H_{SC} + H_{SM-SC}, \label{Htot}
\end{equation}
where $H_{SM}$ and  $H_{SC}$ correspond to the semiconductor and superconductor components, respectively, and the last term characterizes the SM-SC coupling.  Since the SM quantum well is narrow, it is common to work within a single (transverse) band approximation, i.e., to treat the electron gas as a purely 2D system. 

The topological properties of the heterostructure can be captured by a simple BdG effective Hamiltonian that, in the Nambu basis $(\psi_\uparrow,\psi_\downarrow,\psi_\downarrow^\dagger,-\psi_\uparrow^\dagger)^T$, takes the form \cite{Hell2017,Pientka2017}
\begin{eqnarray}
{\cal H}_{\mathrm{eff}}(x,y) &=& \left[-\frac{\hbar^2}{2m^*}(\partial_x^2+\partial_y^2)-\mu\right]\tau_z + i \alpha (\sigma_y \partial_x \!-\!\sigma_x\partial_y)\tau_z \nonumber \\
&+& V_Z(y) \sigma_y + \Delta(y) \tau_+ + \Delta^*(y)\tau_-,             \label{Heff0}
\end{eqnarray}
where $\sigma$ and $\tau$ are Pauli matrices acting on the spin and particle-hole sectors, respectively, where $\tau_{\pm}=(\tau_x\pm i\tau_y)/2$, $m^*$ is the effective electron mass, $\mu$ is the chemical potential, $\alpha$ is the RSOC coefficient, and $V_Z(y)= g(y) \mu_B B/2$ is the Zeeman splitting. The Land\'e factor $g(y)$ is assumed to have different values in the proximitized and unproximitized regions. For simplicity, one can assume $g=0$ underneath the superconductors. The proximity-induced pairing potential is nonzero in the proximitized regions, with $\Delta(y) =  \Delta~\! e^{\pm i \phi/2}$  when $y$ is within the right (left) SC region and $\Delta=0$ otherwise. While most of the physics here is similar to the Majorana nanowires (e.g., RSOC, spin splitting induced by an applied magnetic field, a single subband), the existence of the phase in the pairing potential is a new feature arising from the fact that the SM region here serves as the junction between two superconducting segments as is clear from Fig.~\ref{Fig_V1}.

In general, the effective Hamiltonian given in Eq.~(\ref{Heff0}) has an antiunitary particle-hole symmetry ${\cal P}=\tau_y\sigma_y {\cal K}$, thus belonging to symmetry class D. Since the system is effectively one-dimensional (i.e., the predicted MZMs are bound states emerging within and around the quasi 1D junction region of width $W_J$ smaller than the induced coherence length, at the ends of a finite length junction), 
it has a $\mathbb{Z}_2$ topological classification. However, in the absence of disorder the system can additionally have a (left-right) reflection symmetry $M_y$, which results in an effective ``time-reversal'' symmetry, $\widetilde{\cal T} = \sigma_z M_y {\cal K}$, putting the Hamiltonian in symmetry class BDI with a $\mathbb{Z}$ invariant. This apparent BDI symmetry is, however, impractical since disorder is invariably present in the system, suppressing the ``time-reversal'' symmetry.
Based on the properties of the (simplified) effective Hamiltonian, it was predicted theoretically \cite{Hell2017,Pientka2017} that planar JJ structures with no phase bias ($\phi=0$) can support topological superconducting phases over a wide range of chemical potentials at finite values of the Zeeman field, while tuning the SC phase to $\phi=\pi$ results in a significant expansion of the topological region, even down to $V_Z=0$. Again, the existence of a strong RSOC is the key physics here.  In the absence of RSOC, the system is simply a trivial JJ with no interesting topological properties.

A convenient approach to modeling the hybrid SM-SC structure is to define the problem on a lattice,  express $H$ from Eq.~(\ref{Htot}) as an effective tight-binding Hamiltonian, and integrate out the degrees of freedom associated with the parent SC. The low-energy physics is described in terms of the effective Green's function $G_{SM}$ of the semiconductor, with the (proximity-induced) SC contribution  incorporated as a self-energy \cite{stanescu2017proximityinduced},
\begin{equation}
G_{SM}(\omega) = \left[\omega - {H}_{SM} -\Sigma_{SC}(\omega) \right]^{-1}.    \label{Gwk}
\end{equation}
Here, ${H}_{SM}$ is the semiconductor BdG Hamiltonian, which includes hopping terms, Zeeman contributions associated with the applied magnetic field, as well as  Rashba and Dresselhaus SOC terms corresponding to the discretized (lattice) BdG version of the Hamiltonian \cite{Pakizer2021,Pekerten2022} 
\begin{eqnarray}
H_{SOC} &=& \alpha (k_y\sigma_x- k_x \sigma_y) + \beta \left[(k_x\sigma_x - k_y\sigma_y) \cos \theta \right.  \nonumber \\
&-& \left. (k_x\sigma_y + k_y\sigma_x) \sin\theta \right],   \label{Hsoc}
\end{eqnarray}
where ${\bm k}=(k_x, k_y)$ is the wave vector, $\alpha$ and $\beta$ are the Rashba and Dresselhaus coefficients, respectively, and $\theta$ is the angle between the $[100]$ crystalline direction and the $y$-axis (i.e., the direction perpendicular to the junction, see Fig.~\ref{Fig_V1})~\cite{Pakizer2021,Pakizer2021a,Pekerten2022}. We mention in passing that the analysis presented here could in principle also be extended to planar JJs based on 2D \emph{hole} gases (in, e.g., Ge), which can theoretically host MZMs as well, but for which the SOC takes a more complicated form with terms that are cubic in momentum~\cite{Luethi2023}. 

The self-energy contribution in Eq.~(\ref{Gwk}) is given by the Green's function of the SC film at the SM-SC interface multiplied by the square of the hopping across the interface and contains matrix elements of the form \cite{stanescu2017proximityinduced}
\begin{equation}
\left[\Sigma_{SC}(\omega)\right]_{{\bm i} {\bm j}} = -\frac{\gamma_{{\bm i} {\bm j}}} {\sqrt{\Delta_0^2-\omega^2}}\left(\omega~\!\sigma_0\tau_0 + \Delta_0~\!\sigma_y\tau_y \right), \label{SigSC}
\end{equation}
where ${\bm i}=(i_x, i_y)$ and ${\bm j}=(j_x, j_y)$ label lattice sites within the proximitized regions, $\sigma_\mu$ and $\tau_\nu$ are Pauli matrices associated with the spin and particle-hole degrees of freedom, respectively, and $\Delta_0$ is the pairing potential of the parent SC. We note that a robust proximity effect induced by a thin SC film requires the presence of disorder in the superconductor (e.g., surface roughness), which leads to a quasi-local effective coupling $\gamma_{{\bm i} {\bm j}}$ that decays rapidly with the distance $|{\bm i}-{\bm j}|$ \cite{stanescu2017proximityinduced}. Therefore, the local approximation $\gamma_{{\bm i} {\bm j}}\approx \gamma_{{\bm i} {\bm i}} \delta_{{\bm i} {\bm j}}$, where  $\gamma_{{\bm i} {\bm i}}$ is a strongly position-dependent quantity, represents a natural way to simplify the problem. Furthermore, if the effective SM-SC coupling is weak, the proximity-induced disorder effects are negligible \cite{stanescu2017proximityinduced} and one may consider the uniform coupling approximation $\gamma_{{\bm i} {\bm i}} \approx \langle \gamma \rangle \equiv \gamma$, where $\langle \gamma \rangle = \langle \gamma_{{\bm i} {\bm i}} \rangle_{\bm i}$ is the average effective coupling. 

Additional simplifications in describing the low-energy physics of the hybrid structure can be obtained by using the so-called  ``static approximation'' corresponding to $\sqrt{\Delta_0^2-\omega^2}\approx \Delta_0$ in Eq.~(\ref{SigSC}). Within this approximation, one can define an effective BdG Hamiltonian having spin and particle-hole blocks of the form
\begin{equation}
\left[{H}_{\mathrm{eff}}\right]_{{\bm i} {\bm j}} \!\!=\!\!\left\{
\begin{array}{l}
~~~\!\!\left[{H}_{SM}\right]_{{\bm i} {\bm j}} ; ~~~~~~~~~~~~~~~~~~~~~ {{\bm i}, {\bm j}}  \in\!SM, \\
\!Z^{}\left[{H}_{SM}\right]_{{\bm i} {\bm j}} \!-\!\Delta~\!\sigma_y\tau_y\delta_{j j^\prime}; ~~~~~~{{\bm i}, {\bm j}}  \in\!SC, \\
\!Z^\frac{1}{2}\left[{H}_{SM}\right]_{{\bm i} {\bm j}} ; ~~~~{\bm i} ( {\bm j})\in \!SM, ~ {\bm j}({\bm i})\in\!SC, \label{Heff}
\end{array}\right.
\end{equation}
where $\Delta = \gamma \Delta_0/(\gamma + \Delta_0)e^{\pm i\phi/2}$ is the induced pairing potential within the proximitized ($SC$) regions, $Z=\Delta_0/(\Delta_0+\gamma)$ is the quasiparticle residue, which corresponds to the weight of a low-energy state within the semiconductor component of the heterostructure, while ${{\bm i}, {\bm j}}    \in SM$ and 
${{\bm i}, {\bm j}} \in SC$ label sites within the unproximitized ($SM$) and proximitized ($SC$) regions, respectively. The construction of the effective Hamiltonian can be generalized for multi-orbital systems with SC disorder, when $Z$ and $\Delta$ become orbital- and position-dependent matrices~\cite{stanescu2013majorana}. 
Note that the effective Hamiltonian in Eq.~(\ref{Heff}) has a similar structure as the simplified continuum model in Eq.~(\ref{Heff0}), except that the system parameters, i.e., hopping, SOC coefficients, and Zeeman field 
%and disorder potential, 
within the $SC$ regions, as well as the nearest-neighbor contributions that couple the $SC$ and $SM$ regions, are renormalized. In the strong coupling regime, $\langle \gamma \rangle > \Delta_0$, this 
proximity-induced low-energy renormalization can be significant. The low-energy spectrum obtained by diagonalizing the effective Hamiltonian is accurate for energies much smaller 
than the parent SC gap, $\Delta_0$, and the approximation becomes exact at zero energy. Thus, the topological properties of the system can be obtained from ${H}_{\mathrm{eff}}$. In particular, 
the topological phase diagram of an infinite, uniform Josephson junction structure can be obtained by calculating the ${\mathbb Z}_2$ topological invariant (the so-called topological 
charge)~\cite{kitaev2001unpaired,schnyder2008classification,Ghosh2010,tewari2012topologicala}
\begin{equation}
Q={\rm sign} \left\{{\rm Pf}[H_{\mathrm{eff}}(0)U_P]~\! {\rm Pf}[H_{\mathrm{eff}}(\pi/a)U_P]\right\}, \label{QPf}
\end{equation}
with $Q=+1$ and $Q=-1$ corresponding to the trivial and topological phases, respectively. In Eq. (\ref{QPf}) ${\rm Pf}[\dots]$ designates the Pfaffian and $H_{\mathrm{eff}}(k_x)U_P$, with ${\cal P}=U_P {\cal K}$ representing the particle-hole operator, is a skew-symmetric matrix. Since ${\rm Pf}[A^2]={\rm Det}[A]$ and $|{\rm Det}[U_P]|=1$, a TQPT, which corresponds to the vanishing of one of the Pfaffians in Eq.~(\ref{QPf}), implies ${\rm Det}[H_{\mathrm{eff}}(k_x)]=0$ at $k_x=0$ or $k_x=\pi/a$. As the system is characterized by a large gap at $k_x=\pi/a$ over the entire (relevant) parameter space, this implies that the TQPT is associated with the vanishing of the quasiparticle gap at $k_x=0$.

\subsection{Geometric effects in planar Josephson junction structures} \label{S5_2}

In this subsection we consider the impact of the ``effective geometry'' of the planar device on the emergence and stability of the topological superconducting phase. Here, by ``effective geometry'' we mean the actual shape and size of the superconducting and (unproximitized) semiconductor regions, as well as the effects of applied gate potentials, which control the confinement of electrons underneath the SC films and can be used to deplete certain unproximitized regions. As illustrative examples, we focus on three different aspects: (i) the gate-controlled crossover between the nanowire and the Josephson junction regimes in hybrid systems with narrow SC films, (ii) the optimization of the topological gap as function of the SC width, and (iii) effective SOC engineering in spatially modulated planar devices. Details regarding these aspects can be found in Refs.~\cite{Paudel2025,Paudel2021,woods2020enhanced}.

\begin{figure}[t]
\centering
\includegraphics[width=0.48\textwidth]{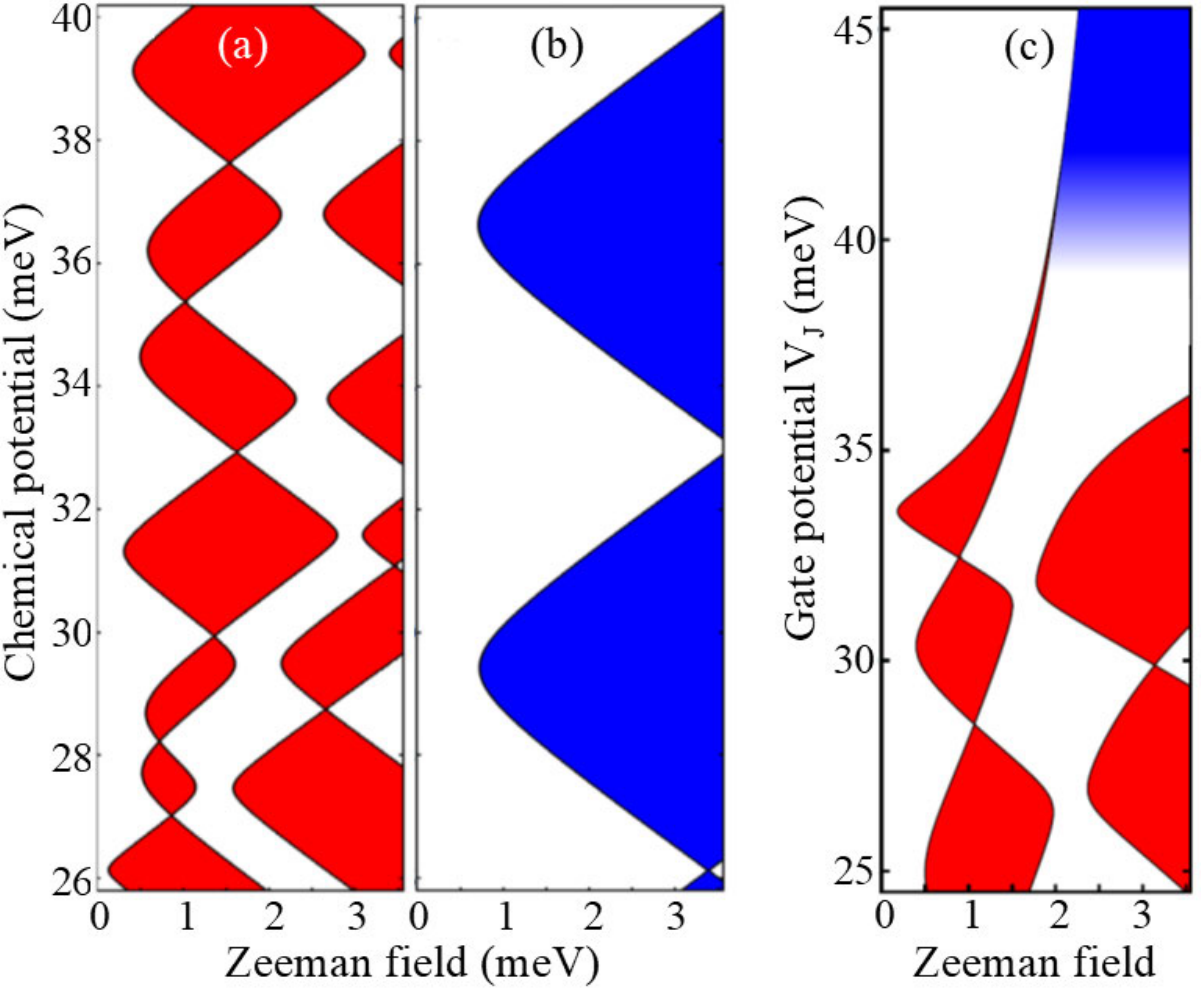}
\caption{(a) Topological phase diagram as a function of Zeeman field and chemical potential for a planar JJ device with  $W_J=90~$nm,  $W_{SC}=150~$nm, $V_J=25~$meV, relative SC phase difference $\phi=0$, Rashba coupling $\alpha\cdot a= 250~$meV$\cdot$\AA, and effective SM-SC coupling $\gamma=0.75~$meV. The white areas are topologically trivial, while the red regions correspond to a topological superconducting phase. The device is in the ``Josephson junction'' regime, $\mu > V_J$. (b) Same structure with an applied gate potential $V_J=45~$meV, i.e., in the ``nanowire'' regime, $\mu<V_J$. The blue areas corresponding to the topological superconducting phase have the typical nanowire outline. (c)  Crossover between the ``nanowire'' and the ``Josephson junction'' regimes for a planar structure with chemical potential $\mu=35~$meV. The “nanowire” topological region (blue) is adiabatically connected to a trivial “Josephson junction” region, while the topological JJ phase (red) collapses into the phase boundaries associated with the “nanowire” regime. (Adapted from Ref.  \cite{Paudel2025}.)}
\label{Fig_V2}
\end{figure}

A typical topological phase diagram for a planar JJ structure with narrow SC films is shown in Fig.~\ref{Fig_V2}(a). The red regions correspond to a topological superconducting phase (i.e., $Q=-1$) and the phase boundaries correspond to the vanishing of the (bulk) quasiparticle gap at $k_x = 0$. Upon increasing the applied gate potential, when $V_J > \mu$, the junction region gets depleted and the structure becomes equivalent to two decoupled nanowires. The corresponding topological phase diagram is shown in Fig.~\ref{Fig_V2}(b). At low field, the topological regions (blue shading) have the characteristic ``hyperbolic'' shape corresponding to the topological condition $V_Z > \sqrt{\gamma^2 + (\mu-E_n)^2}$, where $E_n$ is the energy corresponding to the bottom of the transverse band $n$. To characterize the transition between the ``Josephson junction'' and ``nanowire'' regimes, we fix the chemical potential ($\mu=35~$meV) and vary the potential $V_J$ in the junction region. The corresponding topological phase diagram, shown in Fig.~\ref{Fig_V2}(c), reveals a crossover, with
the topological “nanowire” region (blue) being adiabatically connected to a trivial “Josephson junction” region. Physically, this corresponds to the two pairs of MZMs hosted by the (initially decoupled) wires hybridizing and acquiring a finite energy as $V_J$ is lowered.  Note, on the other hand, that the JJ topological phase (red) requires a finite coupling between the two SC films and collapses when $V_J > \mu$.

\begin{figure}[t]
\centering
\includegraphics[width=0.48\textwidth]{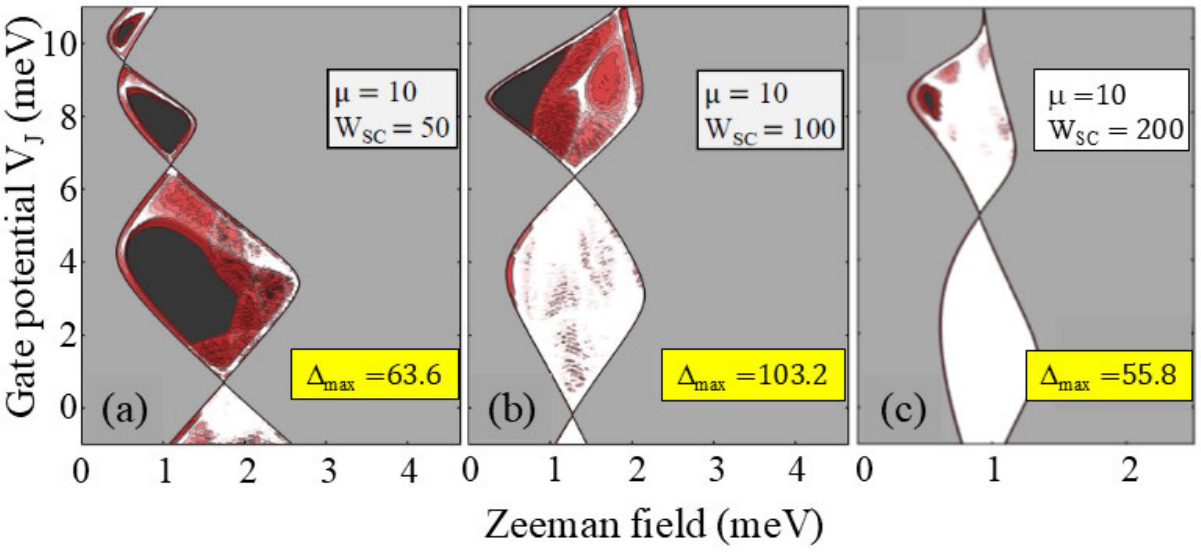}
\caption{Density of states (DOS) at $E^*=35~\muup$eV within the lowest-field topological region for JJ structures with $W_J=90~$nm,  $\phi=0$, $\mu=10~$meV, $\alpha\cdot a= 250~$meV$\cdot$\AA, $\gamma=0.75~$meV, and narrow superconductors ($W_{SC}=50$, $100$, and $200~$nm from left to right). The black regions correspond to zero DOS, i.e., quasiparticle gaps larger than $E^*$. The calculated maximum values of the topological gap, $\Delta_{\mathrm{max}}$, corresponding to each structure are given in $\muup$eV. (Adapted from Ref.~\cite{Paudel2025}.)}
\label{Fig_V3}
\end{figure}

A problem of significant practical importance concerns the optimization of the system parameters to maximize the topological gap, in particular the optimization of the SC films width, $W_{SC}$. (This is very different from the nanowire situation where the topological gap is maximized simply by increasing the RSOC strength and reducing disorder.) The analysis in Ref.~\cite{Paudel2025} shows that the topological gap has a non-monotonic dependence on $W_{SC}$, being  maximized in structures with narrow superconductor films of width ranging between about $100~$nm and $200~$nm, the specific value increasing slightly with the chemical potential (for $10~\mathrm{meV} \leq \mu \leq 40~$meV). The corresponding maximum values of the topological gap can be as high as $40\%$ of the parent superconducting gap, significantly larger than topological gap values characterizing wide-superconductor structures. An illustrative example is shown in Fig.~\ref{Fig_V3}. 

\begin{figure}[t]
\centering
\includegraphics[width=0.42\textwidth]{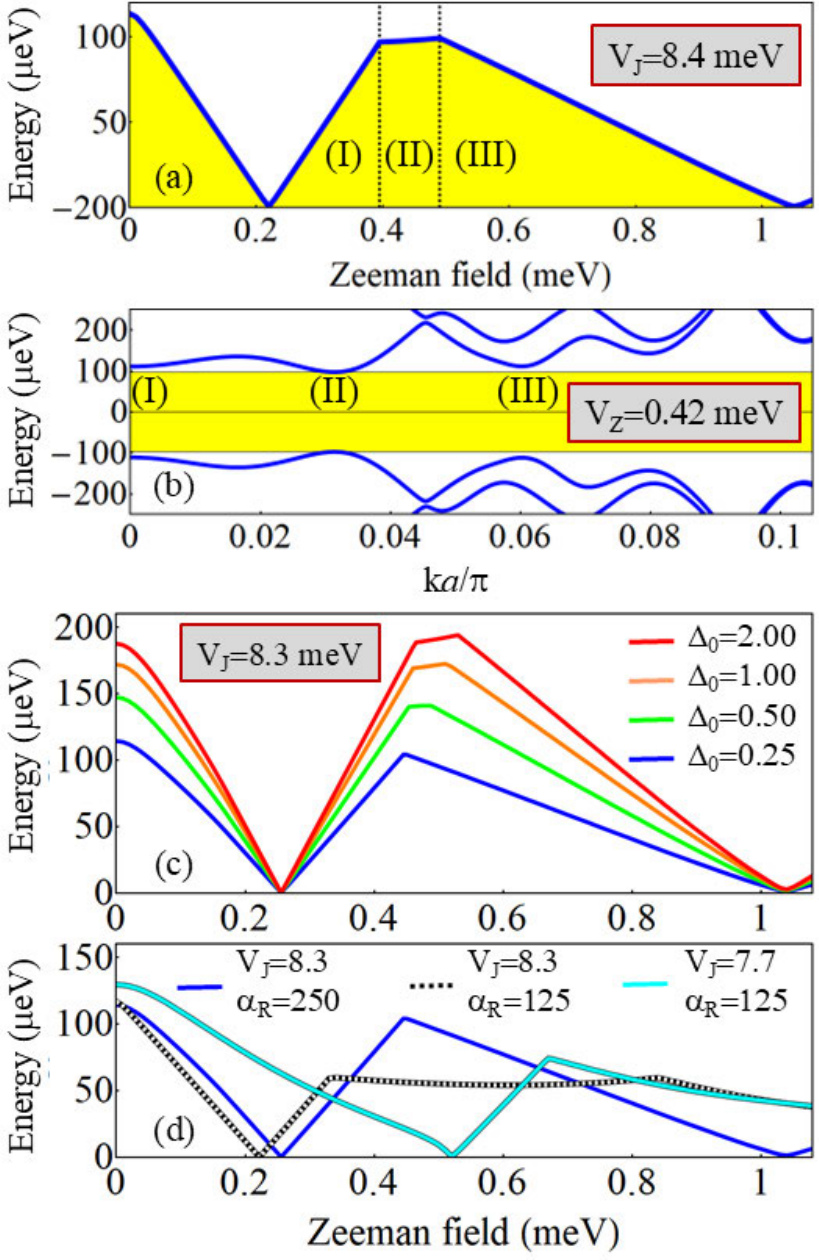}
\caption{(a) Quasiparticle gap as a function of the applied Zeeman field for a JJ structure with system parameters as in Fig.~\ref{Fig_V3}(b) and junction potential slightly above the optimal value $V_J^*=8.3~$meV. The topological gap edge within the regions labeled (I), (II), and (III) is controlled by the corresponding low-$k_x$ modes shown in panel
(b), which illustrates the dependence of the BdG spectrum on the wave vector $k_x$ for $V_Z=0.42~$meV [corresponding to region (II)]. The modes characterized by Fermi wave vectors larger than about $0.08~\!\pi/a$ have larger values of the quasiparticle gap. (c) Quasiparticle gap as a function of the Zeeman field for a JJ structure with different values of the parent SC gap $\Delta_0$ (given in meV), the other system parameters being the same as in (a). (d) Dependence of the quasiparticle gap on the Zeeman field for a JJ structure with reduced Rashba coupling $\alpha_R= \alpha\cdot a$ (given in meV$\cdot$\AA). The blue line corresponds to a cut through the maximum topological gap in Fig.~\ref{Fig_V3}(b). (Adapted from Ref.~\cite{Paudel2025}.)}
\label{Fig_V4}
\end{figure}

For JJ structures with $W_{SC}\gtrsim 100~$nm and $\phi=0$ the large-gap regions correspond to a regime characterized by values of the junction potential $V_J$ comparable to (but smaller than) the chemical potential, i.e., a nearly depleted junction [see, e.g., Figs.~\ref{Fig_V3}(b) and (c)]. In systems with a SC phase difference $\phi=\pi$ the large gap topological region expands significantly, but, remarkably, the maximum topological gap values are comparable to those characterizing the $\phi=0$ regime~\cite{Paudel2021,Paudel2025}. To get further insight into the relevant low-energy physics, we calculate the dependence of the quasiparticle gap on the Zeeman field in the vicinity of the optimal control parameters that maximize the topological gap. A typical example is shown in Fig.~\ref{Fig_V4}(a). The topological gap edge within the regions labeled (I), (II), and (III) is controlled by the corresponding low-$k_x$ modes shown in panel (b). Mode (I), for example, has a minimum at $k_x=0$ and is responsible for the closing and reopening of the gap at the TQPT (corresponding to $V_Z\approx 0.23~$meV). Note that the low-$k_x$ modes, which correspond to the top occupied transverse bands, generically control the low-energy physics within the top topological ``lobe'' (with $V_J$ less than, but comparable to $\mu$), where the large-gap regions are typically located for systems with $\phi=0$ \cite{Paudel2025}. The specific energy values of these modes depend (in addition to the control parameters $V_J$ and $V_Z$) on system parameters such as the width of the SC regions $W_{SC}$, the parent SC gap $\Delta_0$, and the RSOC coefficient $\alpha$. This leads to a nontrivial (typically nonlinear or even nonmonotonic) dependence of the maximum topological gap on the system parameters. 

The nonmonotonic dependence on $W_{SC}$, which enters via the boundary conditions satisfied by the relevant low-$k_x$ modes, has been mentioned above (see Fig.~\ref{Fig_V3}). The dependence on the parent SC gap $\Delta_0$ is illustrated in Fig.~\ref{Fig_V4}(c). Note that the topological phase boundaries (associated with the vanishing of the quasiparticle gap) are independent of $\Delta_0$, being controlled by the effective SM-SC coupling $\gamma$. This property is clearly revealed by Eq.~(\ref{SigSC}), as the zero-energy limit of $\Sigma_{SC}$ (which is the relevant limit at the TQPT) is independent of $\Delta_0$. Also, while the maximum topological gap increases with $\Delta_0$, this enhancement is nonlinear. For example, doubling the size of the original gap (i.e., having $\Delta_0=0.5~$meV) corresponds to an enhancement by about  $50\%$ of $\Delta_{\mathrm{max}}$, while increasing the parent SC gap eight times roughly doubles the size of the maximum topological gap.

Finally, to highlight the key role played by the RSOC, in Fig.~\ref{Fig_V4}(d) we consider the effect of reducing the strength of the Rashba coupling. Starting with parameters corresponding to a cut through the maximum topological gap in  Fig.~\ref{Fig_V3}(b) (blue line), we reduce the Rashba coupling by a factor of two to $\alpha_R = \alpha\cdot a = 125$ meV$\cdot$\AA. The corresponding topological gap (black dashed line) is characterized by a large type-(II) region with a nearly constant value representing about $55-60\%$ of the original maximum gap  ($\Delta_{\mathrm{max}}\approx 103~\muup$eV). For the system with $\alpha_R=125~$meV the maximum topological gap is obtained at a lower value of the gate potential ($V_J\approx 7.7~$meV), when region (II) shrinks to zero (cyan line). We note that the low-$k_x$ modes (I) and (II) correspond to the spin subbands of the top occupied transverse band. Mode (I) has a minimum at $k_x=0$ that disperses almost linearly with $V_Z$. By contrast, mode (II) has a finite Fermi wave vector $k_F^{(II)}$ and a finite characteristic SOC energy $\epsilon_R = \alpha_R k_F^{(II)}$ that controls its ``robustness'' against the spin-polarizing effect of $V_Z$. In other words, the minimum of mode (II) is controlled by the strength of the Rashba coupling, the corresponding gap increasing with increasing $\epsilon_R$. In turn, this results in a nonlinear enhancement of the maximum topological gap.  

\begin{figure}[t]
\centering
\includegraphics[width=0.42\textwidth]{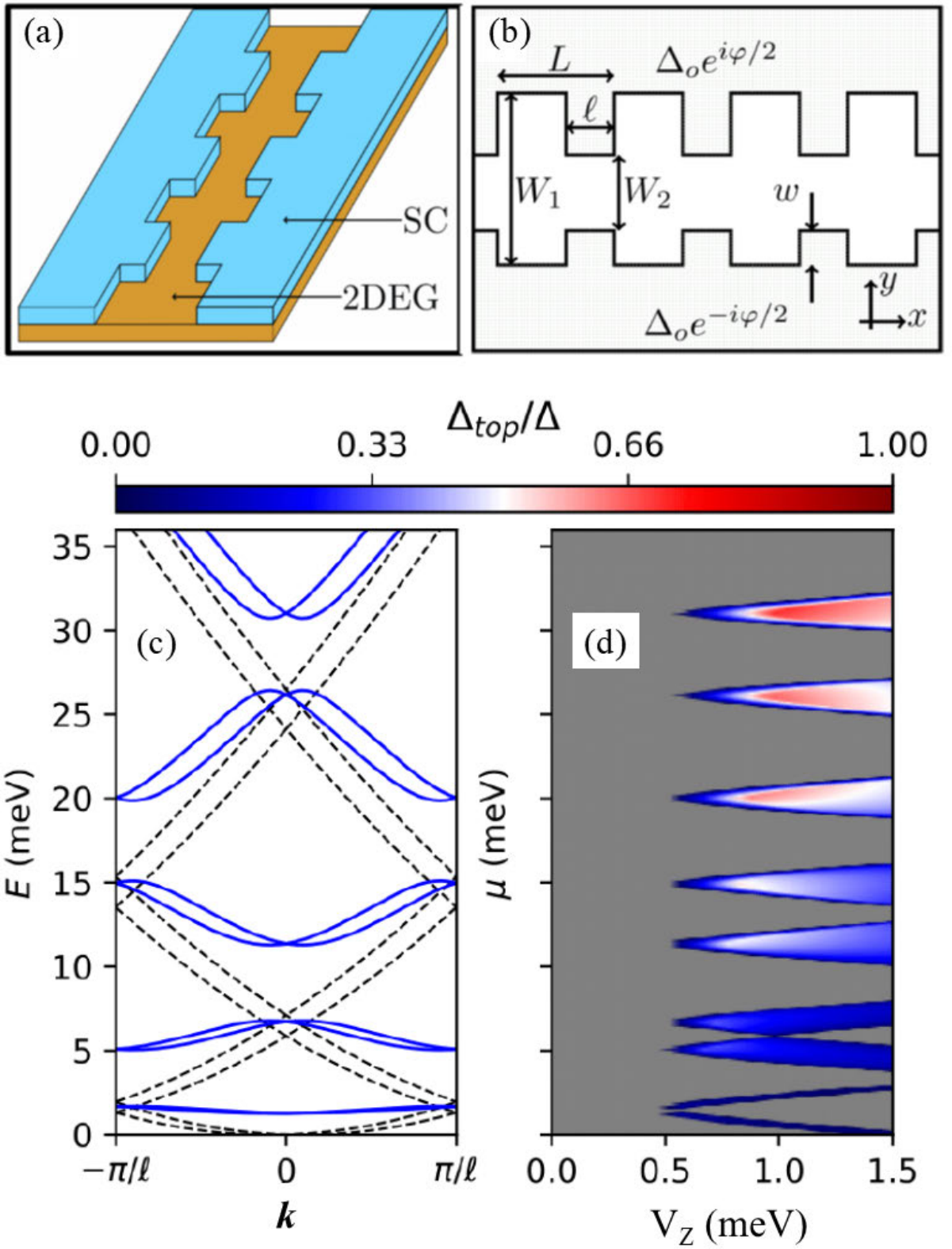}
\caption{(a) Schematic representation of a spatially modulated Josephson junction device. The width of the junction region varies periodically between a maximum value $W_1$ and a minimum value $W_2$. (b) Top view of the junction region that includes four unit cells of the periodic structure. A top gate (not shown) similar to that in  Fig.~\ref{Fig_V1}(a) can be used to control the potential in the junction region.  Note that a spatially modulated Majorana wire can be engineered using a single superconducting strip with periodically modulated width, i.e., a SC having the  shape of the junction region shown above, with the 2DEG outside the SC film being depleted by a top gate potential. (c) Low energy spectrum of a 1D system with RSOC in the presence of a periodic potential (blue solid lines) and the corresponding spectrum of the uniform system (black dashed lines). Note the formation of minibands with increasing SOC splitting (from bottom to top). (d) Topological gap mapped as a function of Zeeman field and chemical potential for a hybrid system with periodic potential. The higher-energy minibands support topological superconducting phases with larger gap values. (Adapted from Refs. \cite{Paudel2021} and \cite{woods2020enhanced}.)}
\label{Fig_V5}
\end{figure}

Considering the key role played by the Rashba SOC in enhancing the topological gap and, consequently, the stability of the topological phase, increasing this coupling represents a problem of major practical importance. Here, we briefly review a less conventional possible solution involving the geometric engineering of the effective RSOC in planar SM-SC devices. The basic idea is to generate a strong effective periodic potential by spatially modulating the structure and then exploit the larger Rashba splitting that characterizes the (higher-energy) minibands induced by the periodic potential. The proposal can be implemented for both nanowires \cite{woods2020enhanced} and Josephson junction devices \cite{Paudel2021} realized in planar SM-SC hybrid structures. To illustrate the idea, we sketch a planar Josephson junction device having a periodically modulated junction width in Figs.~\ref{Fig_V5}(a) and (b). A similar ``Majorana waveguide'' device can be realized in a planar hybrid structure consisting of a 2DEG proximitized by a single SC strip with spatially modulated width (i.e., having a shape similar to that of the junction region in Fig.~\ref{Fig_V5}), with the unproximitized region being depleted by a top gate potential~\cite{woods2020enhanced}. The role of the spatial modulation is to create a strong (effective) periodic potential, which would be difficult to realize using gate potentials due to strong screening. In turn, the effective periodic potential induces minibands within the reduced Brillouin zone, as illustrated in Fig.~\ref{Fig_V5}(c). The higher-energy minibands are characterized by larger values of the SOC splitting, i.e., larger effective Rashba coefficients. Thus, by optimizing the geometry of the periodically modulated structure one can  engineer devices with effective RSOC coefficients much larger than the nominal values for the corresponding uniform structure. 
Finally, the larger effective RSOC results in larger values of the topological gap. In the case of the proposed spatially modulated planar Josephson junction \cite{Paudel2021}, the maximum topological gap was enhanced by a factor $\sim 2$ as compared to its counterpart characterizing the uniform structure, even without optimizing the geometry. 
Additional investigations of the potential benefit of this proposal in the presence of disorder, as well as addressing practical challenges (e.g., regarding the length scales and required precision of the patterning), are necessary future steps. Nonetheless, the possibility of engineering an effective RSOC (significantly different from its nominal value) by periodically patterning planar SM-SC hybrid structures  illustrates  an interesting aspect of the Rashba physics.

\subsection{The superconducting diode effect} \label{S5_4}

We conclude this section with a brief discussion of the {\em superconducting diode effect} (SDE)~\cite{Nadeem2023}---a phenomenon in which SOC plays a prominent role and which could be used as a tool for estimating the SOC strength in planar Josephson junction structures. In essence, the SDE is the superconducting analog of the familiar direction-selective transport in a semiconductor diode with a p–n junction, i.e., a nonreciprocity of the supercurrent associated with  its preferential flow in one direction. The phenomenon was recently observed in noncentrosymmetric superconductors with finite-momentum Cooper pairing, e.g.,  junction-free superconducting lattices \cite{Ando2020,Miyasaka2021,Narita2022,Masuko2022} or thin films \cite{Hou2023}, and Josephson junctions \cite{Baumgartner2022,Santamaria2022,Jeon2022,Golod2022,Wu2022,DMerida2023,Lotfizadeh2024}. 
There are different mechanisms that can lead to SDE based on the simultaneous breaking of space-inversion and time-reversal symmetry, magnetochiral anisotropy, finite-momentum Cooper pairing, and spin-orbit coupling.  A recent review of these mechanisms, as well as materials considerations and a perspective on future directions, can be found in Refs.~\cite{Nadeem2023,Ma2025}. Here, we sketch the basic theoretical elements that highlight the role of SOC in the emergence of the SDE in hybrid SM-SC Josephson junction structures. 

The efficiency of the diode effect can be parametrized in terms of the asymmetry of the forward and reverse critical currents $I_c^+$ and $I_c^-$, 
\begin{equation}
\eta = \frac{I_c^+ - I_c^-}{I_c^+ + I_c^-}. 
\end{equation}
Theoretical studies \cite{Daido2022,Yuan2022,He2022,Ilic2022} have shown that, in addition to temperature and system parameters such as the hopping amplitude, magnetic field,  and  chemical potential, $\eta$ depends on the induced Cooper pairing momentum and spin-orbit coupling. For example, in a Josephson junction with Rashba SOC and applied Zeeman field, a nonzero $\eta$ is associated with an unconventional current-phase relation (CPR) having a non-sinusoidal form and a finite phase shift \cite{Yokoyama2014,Bergeret2015}.  Phenomenologically,  the minimal ingredient that captures this behavior is a second-order harmonic term in the CPR, $I \approx I_{c1} \sin(\phi+\phi_0)  +  I_{c2}\sin(2\phi)$. Numerically, 
the critical currents can be determined by calculating the BdG spectrum as a function of the phase difference, $E_n(\phi)$, which gives the free energy
\begin{equation}
F=-2k_B T\sum_{E_n>0}\ln \left[2\cosh\left(\frac{E_n}{2k_B T}\right)\right],
\end{equation} 
where $T$ is the temperature. The CPR can be obtained as 
\begin{equation}
I(\phi) = \frac{2e}{\hbar}\frac{d F}{d \phi}. 
\end{equation}
Finally, the forward (reverse) critical currents are given by the maximum (minimum) of $I(\phi)$ with respect to the phase,
\begin{equation}
I_c^+ =\left\vert\max_{\phi}[I(\phi)]\right\vert, ~~~I_c^- =\left\vert\min_{\phi}[I(\phi)]\right\vert.
\end{equation}

\begin{figure*}[tb]
    \centering
    \includegraphics[width=0.9\textwidth]{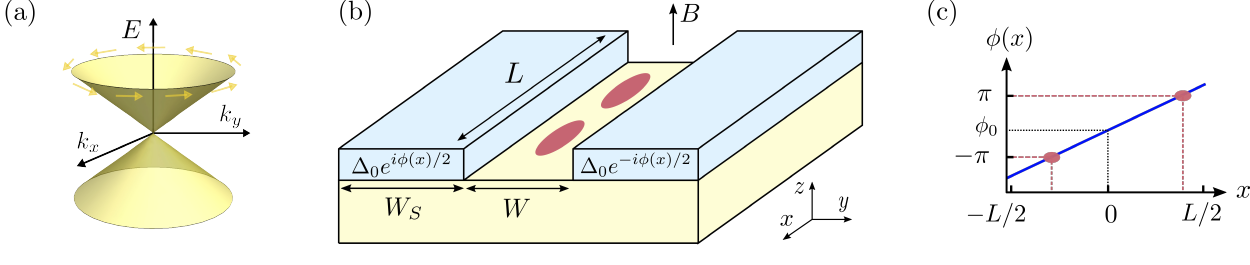}
    \caption{(a) The energy spectrum of a TI surface state consists of a single non-degenerate Dirac cone with spin-momentum locking. (b) A planar Josephson junction formed by two $s$-wave superconductors (blue) placed on the surface of a 3D TI (yellow) in the presence of a perpendicular magnetic field of strength $B$. Due to the magnetic field, the phase difference between the two superconductors varies linearly with the position along the junction [see panel (c)]. Majorana zero modes (red dots) are predicted to emerge at Josephson vortices where the superconducting phase difference is an odd multiple of $\pi$. We assume that the TI is thick enough such that only states near the top surface get proximitized, while the bottom surface remains normal. (Adapted from Ref.~\cite{laubscher2024detection}.)
    }
    \label{fig:ti_vortex}
\end{figure*}

Recent experimental investigations of the SDE in InAs/Al devices similar to those used for studying Majorana physics \cite{Lotfizadeh2024} have reported signatures consistent with a SOC-induced asymmetry of the critical current. More specifically, if the in-plane magnetic field is applied parallel to the current (i.e., along the $y$ direction in Fig.~\ref{Fig_V1}), no signature of nonreciprocity was observed. If, on the other hand, the field is applied perpendicular to the current (i.e., parallel to the junction), a finite difference between the forward and reverse critical currents emerges. This difference changes sign when the in-plane field direction is flipped, $B_x \rightarrow - B_x$. In addition, the magnetic fields at which the critical currents reach their maximum values  are shifted away from zero by a quantity $B_x^*$, which is positive (negative) for  a critical current corresponding to positive (negative) bias. This phenomenology is consistent with RSOC-induced nonreciprocity, as obtained theoretically. Furthermore, by comparing the shift $B_x^*$ of the SDE with the values obtained from numerical simulations, the study estimated the Rashba parameter of the device to $\alpha \approx 100~$meV$\cdot$\AA~\cite{Lotfizadeh2024}. Thus, measurements of the superconducting diode effect in planar SM-SC Josephson junctions provide a useful method for estimating the SOC strength of actual Majorana devices. The SDE is an interesting unexpected manifestation of the Rashba effect in a situation (i.e., superconducting physics) completely distinct from the context where the Rashba effect originated~\cite{rashba1959symmetry,bychkov1984properties}.

%%%%%%%%%%%%%%%%%%%%%%%%%%%%%%%%%%%%%%%%%%%%%%%%%%%%%%%%%%%%%%%%%%%%%%%%%%%%%%%%%%
%%%%%%%%%%%%%%%%%%%%%%%%%%%%%%%%%%%%%%%%%%%%%%%%%%%%%%%%%%%%%%%%%%%%%%%%%%%%%%%%%%

\section{TI-based planar Josephson junctions}
\label{sec:TI_JJ}

Following the seminal works by Fu and Kane~\cite{fu2008superconducting,fu2009jospehson} (see also Sec.~\ref{sec:2d}), various setups have been theoretically proposed to realize Majorana zero modes in TI-based systems. Prominent examples utilize proximitized TI nanowires~\cite{cook2011majorana,cook2012stability}, magnetic vortices in planar SC--TI heterostructures~\cite{hosur2011majorana,ioselevich2011anomalous,chiu2011vortex}, or Josephson vortices in planar SC--TI--SC junctions~\cite{grosfeld2011observing,potter2013anomalous}. In all of these approaches, SOC plays a fundamental (albeit, implicit) underlying role in that it generates the single spin-momentum-locked Fermi surface of the TI surface state [see Fig.~\ref{fig:ti_vortex}(a) for a sketch] that enables a $p$-wave component of the proximity-induced SC pairing even if the bulk superconductor is purely $s$-wave. All TI materials must have strong SO coupling.

Here, we will focus on an approach based on SC--TI--SC Josephson junctions [see Fig.~\ref{fig:ti_vortex}(b) for a schematic setup], the geometry of which resembles the semiconductor-based planar JJs discussed in Sec.~\ref{sec:JJ}. However, since the TI surface state by itself already provides a single non-degenerate Fermi surface, no magnetic field is in principle required to get an effective $p$-wave component of the induced superconducting pairing in this case. Indeed, Fu and Kane showed~\cite{fu2008superconducting} that a planar SC--TI--SC junction with a phase difference of $\pi$ across the junction hosts a Kramers pair of counterpropagating quasi-1D Majorana modes. Localized MZMs can then be obtained if, in addition, a small perpendicular magnetic field is applied to the system, which induces a spatial variation of the gauge-invariant phase difference across the junction, see Fig.~\ref{fig:ti_vortex}(c). In this case, so-called Josephson vortices binding localized MZMs emerge at points where the local phase difference is an odd multiple of $\pi$~\cite{grosfeld2011observing,potter2013anomalous}. These Josephson vortices are so-named because the supercurrent circulates around them just like for a conventional vortex. However, compared to conventional vortices in TI-SC heterostructures, the advantage of the JJ platform is that it readily allows for the creation of an {\emph{array} of MZMs, the number and positions of which can directly be controlled by adjusting the magnetic flux through the junction and the global phase offset between the two superconductors (controllable, e.g., via a flux loop). Furthermore, only very small magnetic fields on the order of mT are required to induce a couple of vortices in junctions of realistic size.

To illustrate in more detail how the vortex MZMs emerge, we model the SC--TI--SC junction by the Bogoliubov-de Gennes (BdG) Hamiltonian $H=\frac{1}{2}\int d\boldsymbol{r}\,\Psi^\dagger\mathcal{H}\Psi$ with $\Psi=(\psi_\uparrow,\psi_\downarrow,\psi_\downarrow^\dagger,-\psi_\uparrow^\dagger)^T$ and
\begin{equation}
\mathcal{H}=( \hbar v\boldsymbol{\sigma}\cdot\boldsymbol{\pi}-\mu)\tau_z+[\Delta(x,y)\tau_++\mathrm{H.c.}],\label{eq:H_cont}
\end{equation} 
where $v$ is the Fermi velocity of the surface state, $\boldsymbol{\sigma}=(\sigma_x,\sigma_y)$ is the vector of Pauli matrices acting in spin space,  $\boldsymbol{\pi}=(\pi_x,\pi_y)=(-i\partial_x-\frac{e}{\hbar} A_x\tau_z,-i\partial_y)$ is the vector of momentum with $A_x$ the vector potential in the Landau gauge, $\mu$ is the chemical potential, $\tau_{x,y,z}$ are Pauli matrices acting in particle-hole space, $\tau_\pm=(\tau_x\pm i\tau_y)/2$, and $\Delta(x,y)$ is the position-dependent proximity-induced superconducting gap. Note that the term $\propto v$ takes a similar form as the two-dimensional RSOC term discussed in Secs.~\ref{sec:2d} and \ref{sec:JJ}. However, the crucial difference is that the TI surface state naturally realizes a single non-degenerate spin-momentum locked Fermi surface even if time-reversal symmetry is intact, whereas, in the absence of time-reversal symmetry breaking, the 2DEGs discussed in the earlier sections will always exhibit two Fermi surfaces.

In our description of the Josephson junction above, we have assumed that the TI is thick enough such that only states near the top surface get proximitized, while the bottom surface remains in the normal state, such that we can consider only a single TI surface state. (Note that the second surface at the bottom must exist in order to satisfy fermion doubling, but it could, in principle, be far from the top surface for a thick enough TI layer.) We also assume that the magnetic field acts only in the normal region of the Josephson junction, such that the vector potential takes the simple form $A_x(y)=BW/2$ for $y<-W/2$, $-By$ for $|y|\leq W/2$, and $-BW/2$ for $y>W/2$, where $W$ is the width of the junction. Furthermore, we neglect the Zeeman splitting induced by the magnetic field as this splitting is expected to be very small for the range of magnetic fields we are interested in ($B\sim$ a couple of mT). The superconducting gap is taken to be
\begin{equation}
\Delta(x,y)= 
\Delta_0 e^{-i\mathrm{sgn}(y)\phi(x)/2}\Theta(|y|-W/2),\label{eq:scgap}
\end{equation}
where $\Delta_0>0$ is real, $\phi(x)$ is the superconducting phase difference between the two superconductors, and $\Theta$ is the Heaviside step function. In the limit where the junction length is much smaller than the Josephson penetration length, the phase becomes a linear function of the position along the junction with a slope set by the magnetic field~\cite{barone1982physics},
\begin{equation}
\phi(x)=\frac{2\pi}{L}\frac{\Phi}{\Phi_0}x+\phi_0,\label{eq:scphase}
\end{equation}
where $\Phi=BWL$ is the magnetic flux piercing the junction, $\Phi_0=h/2e$ is the flux quantum, $L$ is the length of the junction, and $\phi_0$ is the superconducting phase difference at the center of the junction $x=0$.

\begin{figure*}[tb]
    \centering
    \includegraphics[width=1\textwidth]{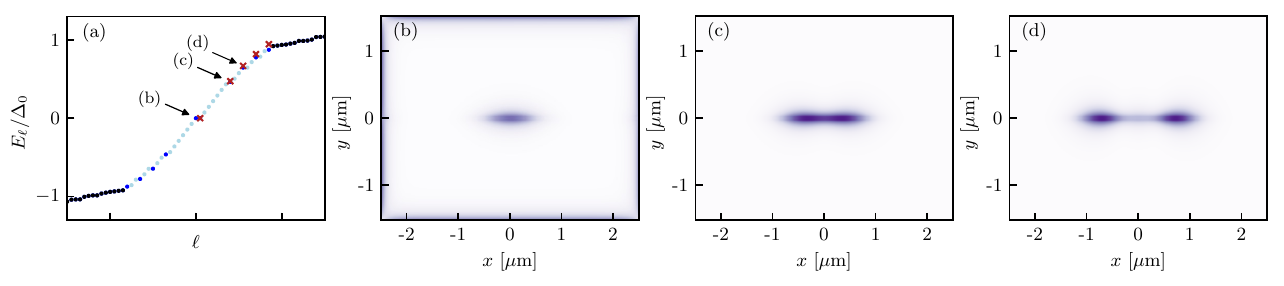}
    \caption{(a) Low-energy BdG spectrum of a discretized version of Eq.~(\ref{eq:H_cont}) (black dots: bulk states, light blue dots: edge states, dark blue dots: CdGM states in the junction) and energies of the CdGM states obtained from the analytical solution given in Eq.~(\ref{eq:energies_1d}) (red crosses; only positive energies shown). The integer $\ell$ enumerates the low-energy eigenstates of the numerical BdG Hamiltonian. (b)-(d) Probability density of the eigenstates labeled by arrows in (a). The state shown in (b) is a superposition of an MZM bound to the center of the vortex and a delocalized MZM at the outer periphery of the superconductors, while the states shown in (c) and (d) are higher-energy CdGM states. The system parameters are $v=4\times 10^5$~m/s, $\Delta_0=1.5$~meV, $\mu=0$, $W=50$~nm, $\Phi=1.3\Phi_0$, $\phi_0=\pi$, $L=5~\mu$m, and $W_S=1.5~\mu$m. (Adapted from Ref.~\cite{laubscher2024detection}.)
    }
    \label{fig:spectrum_ti_vortex}
\end{figure*}

To see how MZMs emerge at the cores of Josephson vortices, it is instructive to start from the gapless Majorana modes that are present if the phase difference is constant $\phi(x)\equiv \pi$. We can then obtain an effective low-energy Hamiltonian by taking the momentum along the junction and a small linear phase gradient into account perturbatively. This leads to an effective low-energy Hamiltonian describing the low-energy physics in the vicinity of an isolated Josephson vortex,
\begin{equation}
\mathcal{H}_{\mathrm{eff}}=-i\hbar \tilde{v}\partial_x \rho_x+m\rho_y,\label{eq:Heff1D}
\end{equation}
where $\rho_{x,y}$ are Pauli matrices acting in the space of the low-energy Majorana modes, $m\approx -\frac{1}{2}\frac{\Delta_0\left(\phi(x)-\pi\right)}{1+W/\xi}$ is an effective mass, and 
%
%\begin{equation}
$\tilde{v}=v\,\frac{[\cos(\mu W/\hbar v)+(\Delta_0/\mu)\sin(\mu W/\hbar v)]}{1+W/\xi}\frac{\Delta_0^2}{\mu^2+\Delta_0^2}$
%\end{equation}
%
is the renormalized velocity.
The Hamiltonian $\mathcal{H}_{\mathrm{eff}}$ can be solved analytically by introducing ladder operators $a,a^\dagger\propto -i\hbar\tilde{v}\partial_x\mp i\lambda\Delta_0 x$ such that $\mathcal{H}_{\mathrm{eff}}\propto\ a\rho_-+a^\dagger\rho_+$. The eigenstates and eigenenergies can then readily be written as 
\begin{align}
|\psi_n\rangle&=\frac{1}{\sqrt{2}}(|n\rangle\otimes|\rho_z=+1\rangle+|n-1\rangle\otimes|\rho_z=-1\rangle),\\
E_n&=\sqrt{2 n\hbar \tilde{v}\lambda\Delta_0},\label{eq:energies_1d}
\end{align}
for $n>0$. Here, we have defined the number eigenstates $|n\rangle$ satisfying $a^\dagger a|n\rangle=n|n\rangle$. The negative energy eigenstates can be obtained by applying the chiral symmetry operator $\rho_z$ so that $|\psi_{-n}\rangle\equiv\rho_z|\psi_n\rangle$ has energy $-E_n$. These in-gap ABSs with energies $\pm E_n$ are so-called Caroli-de Gennes-Matricon (CdGM) states bound to the Josephson vortex. In addition, there is a unique MZM $|\psi_0\rangle=|0\rangle\otimes|\rho_z=+1\rangle$ with energy $E_0=0$. In Fig.~\ref{fig:spectrum_ti_vortex}(a), we plot the first few energy levels $E_n$ from Eq.~(\ref{eq:energies_1d}) as well as the spectrum obtained by numerical exact diagonalization of a discretized version of Eq.~(\ref{eq:H_cont}), showing a good agreement between the analytical and the numerical solution. In Fig.~\ref{fig:spectrum_ti_vortex}(b), we additionally plot the probability density of one of the zero-energy states obtained from the numerical model, showing that there is one isolated MZM located at the center of the vortex as expected from our analytical analysis, while its partner (which is not captured by the analytical solution) is delocalized along the outer periphery of the superconductors. The probability density for the two next CdGM states in the junction is shown in Figs.~\ref{fig:spectrum_ti_vortex}(c) and (d). Note that in addition to the CdGM states, the numerical low-energy spectrum also contains edge states that propagate along the edges of the sample [light blue dots in Fig.~\ref{fig:spectrum_ti_vortex}(a)], which is an artifact of our choice to only model a single 2D surface state of the 3D TI.

While we have focused on a single vortex in the above, it is straightforward to increase the number of vortices by increasing the magnetic flux through the junction. Furthermore, more complicated junction geometries, such as, e.g., trijunctions, could in principle be envisioned, which would allow for braiding of vortices by flux biasing~\cite{hedge2020topological}. Finally, just as for the Rashba 2DEG-based junctions discussed in Sec.~\ref{sec:JJ}, it is interesting to study the supercurrent across the junction and related phenomena such as, e.g., the SDE, which may emerge in more complicated variants of our basic model presented above. In this context it is also worth noting that, while Fraunhofer pattern measurements constitute one of the simplest and most natural ways to experimentally characterize Josephson junctions in a magnetic field, these measurements are generally not well suited to detect vortex MZMs, which do not significantly contribute to the supercurrent density as long as they remain spatially isolated~\cite{laubscher2024detection}.

\section{Conclusions}
\label{sec:conclusions}

We have discussed the key role that Emmanuel Rashba played through the physics of Rashba spin-orbit coupling in the design of artificially engineered semiconductor-superconductor hybrid systems for creating topological superconductivity with Majorana zero modes. The Rashba effect plays two crucial roles: First, it allows superconducting proximity effect from an ordinary $s$-wave metallic superconductor (e.g., Al) to induce effectively `triplet' type $p$-wave superconductivity, which was also anticipated by Rashba in Ref.~\cite{gor2001superconducting}. Second, the MZM-carrying topological superconductivity is enhanced by the RSOC strength with the topological gap increasing with increasing RSOC. Interestingly, however, the TQPT itself is independent of the RSOC strength. If this platform leads to successful topological quantum computing, then the Rashba effect becomes a most important technological tool in the design and optimization not only of non-Abelian anyons, but also of fault tolerant quantum computing. In various personal conversations with one of the coauthors (Sankar Das Sarma), Emmanuel Rashba was always ecstatic about the possibility that this old idea of his, which was neglected by the scientific community for almost 40 years, may lead to the development of disruptive technology (namely, quantum computers), but as is usual for Emmanuel Rashba, he felt that he should get no credit for this idea since he played no role in the theoretical papers of 2009-2010 which led to the semiconductor-superconductor hybrid Majorana platform. Emmanuel Rashba was not only a great physicist, he was also exceptionally modest, and was not comfortable with praise and adulation. His legacy lives, and hopefully, a topological quantum computer using the Rashba effect as the key underlying physical mechanism will make his name immortal both in fundamental science and disruptive technology.

\begin{figure}[tb]
    \centering
    \includegraphics[width=0.85\columnwidth]{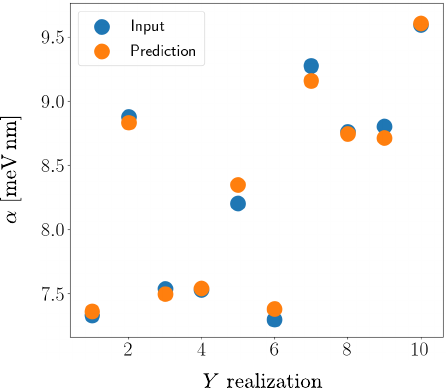}
    \caption{Representative examples of input (blue) and predicted (orange) values for the RSOC strength $\alpha$ obtained from a machine learning approach based on a convolutional neural network. Input values are used in a numerical KWANT simulation to generate the four components of the conductance matrix $G_{\alpha\beta}$ (here $\alpha,\beta\in\{L,R\}$) for an InAs/Al hybrid nanowire connected to two leads in the presence of a specific realization of spatially varying random disorder in the chemical potential. The simulated conductance data, as a function of magnetic field and chemical potential, are then fed into the convolutional neural network, which was previously trained on a separate set of simulated conductance data. The neural network then predicts the value of $\alpha$ based on the input conductance data. We find a remarkable agreement between input and predicted values for $\alpha$ throughout. (Adapted from Ref.~\cite{taylor2024machine}.)
    }
    \label{fig:machine_learning}
\end{figure}

One aspect worth emphasizing is that RSOC may turn out to be the key in the eventual resolution of the main problem preventing experimental progress in the subject, namely, disorder, which strongly suppresses the topological gap, eventually driving the system into a localized phase.  If the RSOC can be enhanced, either by finding suitable new materials or by cleverly engineering the RSOC strength in the current materials, that will go a long way in mitigating disorder effects since larger RSOC would enhance the topological gap, thus suppressing the detrimental disorder effects.  Much more work is necessary searching for new materials with larger RSOC strength as well as trying for ways to engineer larger RSOC strength in existing materials.

Before concluding, we briefly mention and discuss some open questions regarding RSOC in the engineered Majorana platforms. Although the focus of this discussion is the SM-SC nanowire structure, the issues we raise here apply quite generally to all the artificially engineered structures discussed in this article.

The most important open question about RSOC in Majorana nanowires is that the actual strength of the Rashba coupling is quantitatively unknown since these structures are complex multi-layered devices with the SM wire in contact with not only the SC layer, but also dielectric layers and various metallic gates. No direct measurement of the RSOC in situ has been carried out, and in fact, such a measurement is probably quite impossible. One can at best estimate the coupling from indirect measurements (e.g., the induced gap beyond the putative TQPT) using extensive numerical simulations. Since there are many unknown parameters in the problem (e.g., the chemical potential in the SM wire, the precise $g$-factor of the SM, the relevant carrier density, the effective wire length, and most importantly, the unintentional and unknown sample disorder), estimating the unknown RSOC strength is a serious challenge. Since the topological gap is directly proportional to the RSOC strength, a quantitative estimate of the coupling (and specific ideas about how to enhance it through fabrication and engineering) is an extremely important open question. In fact, it is likely that the RSOC strength varies along the wire (leading perhaps to a spatially inhomogeneous gap along the wire) since the inversion asymmetry very likely varies spatially along the wire. These are difficult issues to tackle either experimentally or theoretically.

There has been some recent progress in the estimation of the RSOC strength by using unsupervised deep learning techniques and vision transformer methods where realistic simulated data are used for training the neural networks~\cite{taylor2024machine}. The machine learning simulations seem to be remarkably stable in producing the estimate for the RSOC with very high fidelity, but whether such techniques can be used on real measured data to estimate the coupling strength is unknown. In Fig.~\ref{fig:machine_learning} we reproduce the excellent simulated results for the RSOC strength as obtained in this recent theoretical work, and we refer to the original work for more details on the methodology used to obtain these machine learning estimates of the RSOC strength in the InAs/Al SM-SC hybrid system. It is important to adapt this machine learning approach in the experimental work~\cite{microsoftquantum2023inasal,aghaee2025interferometric} so that the RSOC strength may be estimated more accurately.

In this context, it may also be important to mention another potential open question regarding the nanowire SO coupling. It is known that in the presence of RSOC, the well-known ballistic conductance quantization of 1D nanowires is modified (particularly in the presence of an applied magnetic field) since the RSOC (and the spin splitting) considerably modifies the 1D subband structure---in particular, a helical spin gap opens up for low carrier density---as is clear from Fig.~\ref{fig:setup}(b) in the Introduction. Therefore, the 1D conductance quantization should manifest some subtle structures with increasing characteristic nonmonotonicity (arising from the helical spin gap induced by RSOC and spin splitting) which is absent if there is no RSOC~\cite{pershin2004effect,quay2010observation}. These RSOC-induced features in the quantized conductance have, however, not been observed experimentally although the ballistic conductance quantization itself has been reported in InSb nanowires used in the SM-SC hybrid platforms~\cite{kammhuber2016conductance,qu2016quantized}. The absence of any direct evidence for the RSOC-induced helicity in the conductance quantization experiments remains an important open question in the subject.

\section*{Acknowledgments}

This work is supported by the Laboratory for Physical Sciences through the Condensed Matter Theory Center.
H.P. is supported by US-ONR grant No.~N00014-23-1-2357.

\bibliography{Majorana,Majorana_HP,Majorana_TS}

\end{document}